\documentclass[%
 reprint,
 amsmath,amssymb,
 aps,
]{revtex4-1}

\usepackage{graphicx}
\usepackage{dcolumn}
\usepackage{bm}

\begin{document}


\title{On the possibility of neutrino flavor identification at the highest energies}

\author{A. D. Supanitsky}
\affiliation{Instituto de Astronom\'ia y F\'isica del Espacio (IAFE), CONICET-UBA, Argentina.}
\email{supanitsky@iafe.uba.ar}

\author{G. Medina-Tanco}%
\affiliation{Departamento de F\'isica de Altas Energ\'ias, Instituto de Ciencias Nucleares, Universidad
Nacional Aut\'onoma de M\'exico, A. P. 70-543, 04510, M\'exico, D. F., M\'exico.}%
\email{gmtanco@nucleares.unam.mx}

\date{\today}

\begin{abstract}

High energy astrophysical neutrinos carry relevant information about the origin and propagation of cosmic rays.
They can be created as a by-product of the interactions of cosmic rays in the sources and during propagation 
of these high energy particles through the intergalactic medium. The determination of flavor composition in this 
high energy flux is important because it presents a unique chance to probe our understanding of neutrino flavor 
oscillations at gamma factors $>10^{21}$. In this work we develop a new statistical technique 
to study the flavor composition of the incident neutrino flux, which is based on the multipeak structure 
of the longitudinal profiles of very deep electron and tau neutrino horizontal air showers. Although these 
longitudinal profiles can be observed by means of fluorescence telescopes placed over the Earth's surface, orbital 
detectors are more suitable for neutrino observations owing to their much larger aperture. Therefore, we focus on 
the high energy region of the neutrino spectrum relevant for observations with orbital detectors like the planned 
JEM-EUSO telescope. 
 
\end{abstract}

\maketitle


\section{Introduction}

The existence of neutrinos of $E_\nu \gtrsim 10^{12}$ eV is strongly motivated by the observation of cosmic 
rays up to energies of order of $10^{20}$ eV. These cosmic ray particles, mainly composed by protons and heavier 
nuclei, interact in the sources (see, for instance, Ref. \cite{Ostapchenko:08}) or during their propagation through 
the intergalactic medium \cite{Berezinsky:69}, producing weakly decaying particles such as pions and kaons that 
decay into neutrinos. High energy neutrinos can also be generated as the main product of the decay of superheavy 
relic particles \cite{Aloisio:04,Battacharjee:00}. 

In general, the flavor content of the flux of astrophysical neutrinos is a function of the neutrino energy 
\cite{Kashti:05,Lipari:07,Hummer:11}. One of the main mechanisms that causes the flavor ratio to change is based 
on the fact that pions and muons produced in the sources lose their energy because of the interaction with the 
ambient magnetic field. The pion decay time is shorter than the corresponding one to the muon, and then, at 
sufficiently high energies, the probability of pion decay, prior to significant energy loss, is larger than the 
corresponding one to the muon. As a consequence, the flavor ratio at the source changes from 
$\Phi_{\nu_e}:\Phi_{\nu_\mu}:\Phi_{\nu_\tau} = 1:2:0$ to $\Phi_{\nu_e}:\Phi_{\nu_\mu}:\Phi_{\nu_\tau} = 0:1:0$. 
The transition between theses two regimes depends on the characteristics of the source. Therefore, the determination 
of the flavor content of the flux is of great importance for the understanding of the physical processes that take 
place in the source. The flavor information can also be used to study the more fundamental properties of neutrinos 
and to identify physics beyond the standard model (see, for instance, Refs. \cite{Pakvasa:08,Blenow:09,Beacom:05}). 

Cosmic ray observatories are sensitive to high energy neutrinos. In particular, there are two different
ways to detect them. The first one consists in observing the development of horizontal air showers produced by 
the interactions of the neutrinos with the nucleons of the molecules of the Earth's atmosphere \cite{Berezinsky:75}.
The second one consists in observing the showers produced by the decay of taus generated by the interaction 
of tau neutrinos that propagate through the interior of the Earth \cite{Fargion:99,Fargion:02,Fargion:04}.

Horizontal and quasihorizontal neutrino showers as well as Earth skimming tau neutrino showers can be observed 
by using surface detectors or fluorescence telescopes placed over the Earth's surface or in space. At present no 
astrophysical neutrino has been observed in cosmic ray observatories and then, upper limits on its flux 
have been obtained \cite{chicago:12}.       

In this work we develop a new technique to discriminate between scenarios with different compositions of 
electron and tau neutrinos, by using very deep tau and electron neutrino horizontal showers. This new technique 
is based on the morphology of the longitudinal profiles of the horizontal neutrino showers that can be observed 
by means of fluorescence telescopes. In particular, we study in detail the highest energy region of the neutrino 
flux that is relevant for space observations with the upcoming orbital fluorescence telescopes like JEM-EUSO 
\cite{Takahashi:09} and the proposed Super-EUSO \cite{Petrolini:11} experiment.

\section{Neutrino showers}

Neutrinos can initiate atmospheric air showers when they interact with the nucleons of the air molecules.
The probability that a neutrino interacts in the atmosphere increases with the zenith angle because of the
increase of the number of target nucleons. High energy neutrinos, propagating through the atmosphere,
can suffer charge (CC) and neutral (NC) current interactions. The CC interactions are the most important
for the space observations because in the NC interactions most of the energy is taken by a secondary
neutrino that could produce an observable air shower just in case it suffers a subsequent CC
interaction. The observation of the showers produced by the hadronic component resultant from the NC 
interaction depends on the energy threshold of the telescope. In this work just the showers initiated by 
CC interactions are considered.

As a result of a CC interaction, a very high energy lepton, which takes most of the energy of the incident
neutrino, is generated,
\begin{equation}
\nu_l + N \rightarrow l + X,
\end{equation}
where $N$ is a nucleon, $l=\{ e^-,\mu^-,\tau^- \}$, and $X$ is the hadronic component. Typically, the 
produced lepton $l$ takes $\sim 80\%$ of the neutrino energy at $E_{\nu} \cong 10^{20}$ eV, and the rest 
of the energy goes into the hadronic component $X$. In the case of electron neutrino showers, the produced 
electron together with the hadronic component initiate an air shower right after the neutrino interaction. 
Because most of the times the electron takes most of the neutrino energy, the electron neutrino showers
have characteristics more similar to the electromagnetic showers than to the hadronic showers; however, the 
effects of the hadronic component are not negligible specially for showers developing in dense regions of 
the atmosphere (see Ref. \cite{Supanitsky:11} for details).   

In the case of tau neutrinos, while the hadronic component $X$ initiates a low energy shower immediately 
after the CC interaction (first bang), the tau lepton produced propagates through the atmosphere almost 
without interacting and then, after a given distance, decays. The particles produced in the decay initiate 
a more energetic shower than the first one, producing a second bang. There are several channels for tau 
decay which can be classified in three groups \cite{tauola},
\begin{itemize}

\item Electromagnetic channel: $\tau \rightarrow \nu_\tau+e^-+\nu_{e^-}$.

\item Hadronic channel: $\tau \rightarrow \nu_\tau+X$, where $X$ can be pions, kaons, etc.

\item Muonic channels: $\tau \rightarrow \nu_\tau+\mu^-+\nu_\mu$.

\end{itemize}

The branching ratios of the electromagnetic, hadronic, and muonic channels are $b_{em}=0.18$, $b_{h}=0.645$, 
and $b_\mu = 0.175$, respectively. The tau neutrino showers corresponding to the muonic channel are difficult 
to observe because of the large decay length of the muons at the energies considered, and then, only the electromagnetic 
and hadronic channels are relevant for the detection of tau showers. Therefore, considering just the electromagnetic 
and hadronic channels, $\sim 78\%$ of the tau neutrino showers are of hadronic origin whereas $\sim22\%$ are of
electromagnetic origin.   

Muon neutrino showers are initiated by the electrons generated in the dominant decay channel of the muons 
produced in the CC interaction. However, these showers are difficult to observe because of the large decay 
length of the muon at such high energies, $\lambda(E_\mu) \cong 6.23 \times 10^{10} (E_\mu/10^{19} \textrm{eV})$ km. 
That is why muon neutrino showers are not considered in the subsequent analyses.

Electron neutrino showers are simulated following the procedure introduced in Ref. \cite{Supanitsky:11}, 
and the charge current neutrino-nucleon interaction is simulated by using PYTHIA \cite{pythia}, linked with 
the LHAPDF library \cite{lhapdf}, which allows the use of different sets of parton distribution functions 
(PDFs). The CTEQ6 \cite{cteq6} set of PDFs, the most commonly used at the highest energies, is used in this 
work. Note that the use of a different set of PDFs has a small impact on the simulated neutrino showers 
\cite{Supanitsky:11,AugerNu:11}. The secondary particles, produced in the interaction, are used as input in 
the program CONEX \cite{conex} (v2r2.3), in order to simulate the shower development. The high energy hadronic 
interaction model used for the shower simulations is QGSJET-II \cite{QGSJETII}. However, our main results are 
not strongly dependent on this assumption.

The CC tau neutrino-nucleon interaction and the production of the corresponding tau lepton, are simulated using 
PYTHIA with the CETQ6 set of PDFs. The decay of the tau is simulated with the program TAUOLA \cite{tauola}. Then, 
the particles produced in the decays are used as input in CONEX (with QGSJET-II) in order to simulate the shower 
development. 

A library of horizontal neutrino showers is generated by using the simulation chain described above. The primary 
energy of the neutrinos ranges from $10^{18}$ eV to $10^{20.5}$ eV in steps of $\Delta \log(E_\nu/\textrm{eV}) = 0.25$.
The interaction point of the electron neutrinos and the decay of the taus are just above the Earth's surface, 
the densest region of the atmosphere. Note that the injection point is such that the trajectory of the showers 
starts at the vertical axis of a nadir-pointing orbital telescope at sea level. Therefore, hereafter we denote very 
deep horizontal (VDH) showers to the showers initiated by the interaction of horizontal electron neutrinos or the 
decay of tau leptons (generated by the interaction of horizontal tau neutrinos), such that the interaction point 
of the electron neutrinos, or the decay of the tau leptons, is placed at the vertical axis of a nadir-pointing telescope 
at sea level.

\begin{figure}[!ht]
\includegraphics[width=8cm]{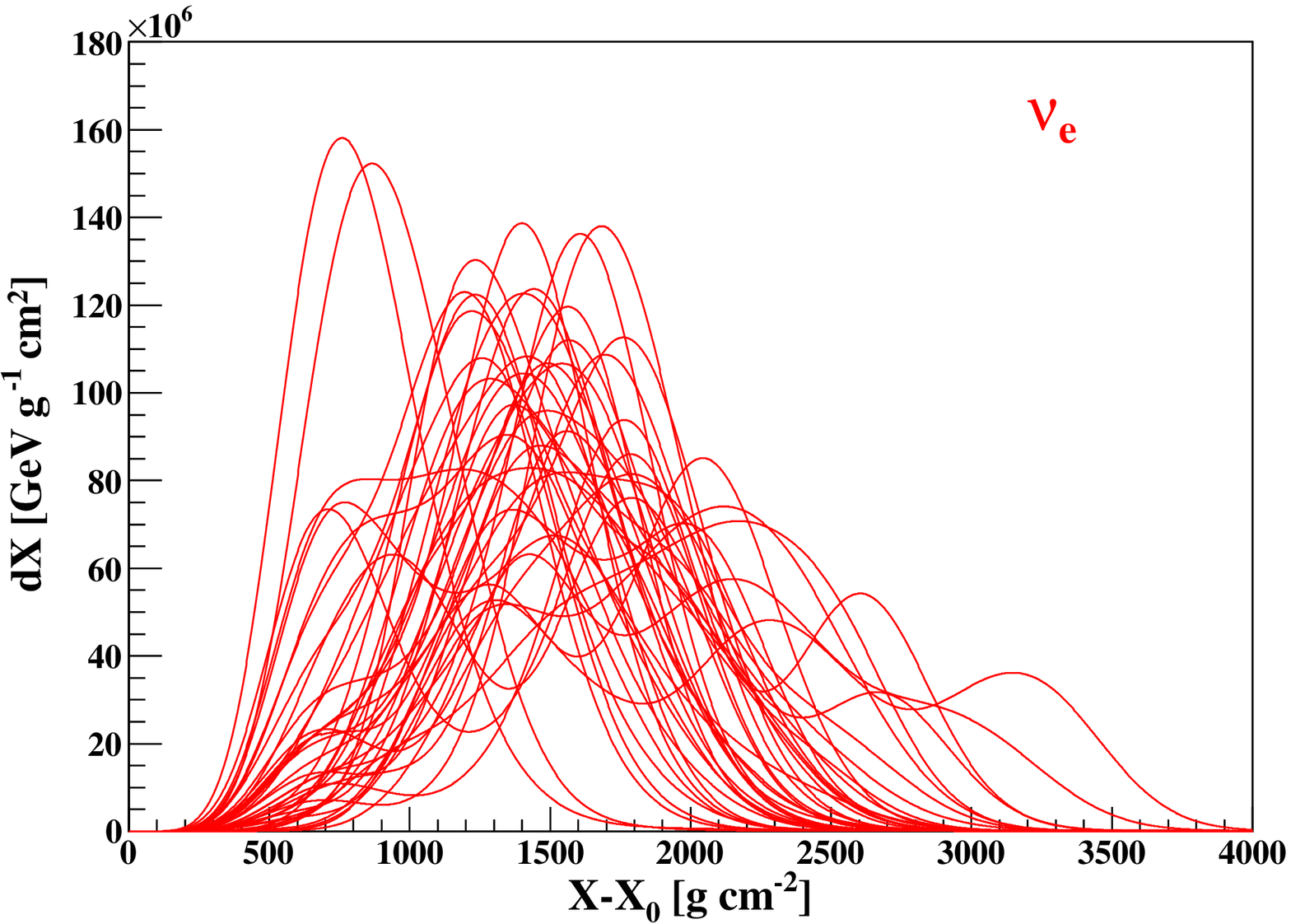}
\includegraphics[width=8cm]{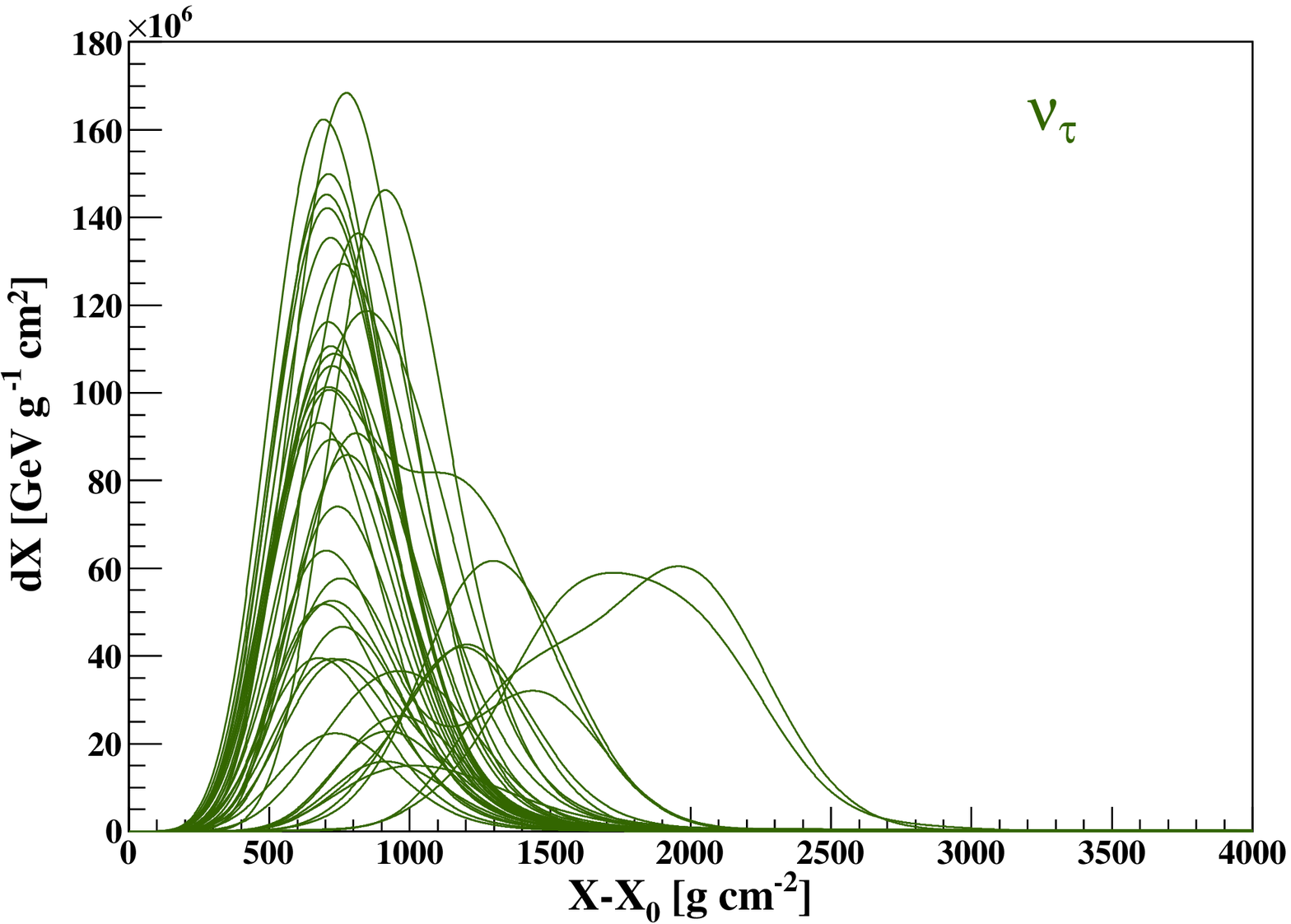}
\caption{\label{Prof} Longitudinal profiles for VDH electron and tau neutrino showers of $E_\nu = 10^{20}$ eV.}
\end{figure}

Figure \ref{Prof} shows the longitudinal profiles (deposited energy per unit of atmospheric depth) for VDH electron 
and tau neutrino showers of $E_\nu = 10^{20}$ eV. The multiple peak structure present in the profiles is caused by 
the fluctuations introduced by the Landau Pomeranchuk Migdal (LPM) effect, which is important for showers dominated 
by the electromagnetic component. Note that, as can be seen from the figure, the VDH electron neutrino showers are 
more affected by the LPM effect.     

\section{Flavor discrimination}
\label{Disc}

As mentioned before, VDH neutrino showers present a multipeak structure caused by the fluctuations introduced by 
the LPM effect. In particular, electron neutrino showers are more affected by this effect because they are dominated 
by the electromagnetic component. In contrast, tau showers are less affected by this effect, because $\sim 78\%$ of 
the times the tau decays are of the hadronic type. Therefore, the number of peaks present in a given longitudinal profile 
should be a good parameter to discriminate between VDH electron and tau neutrino showers.   

\begin{figure}[!h]
\includegraphics[width=8cm]{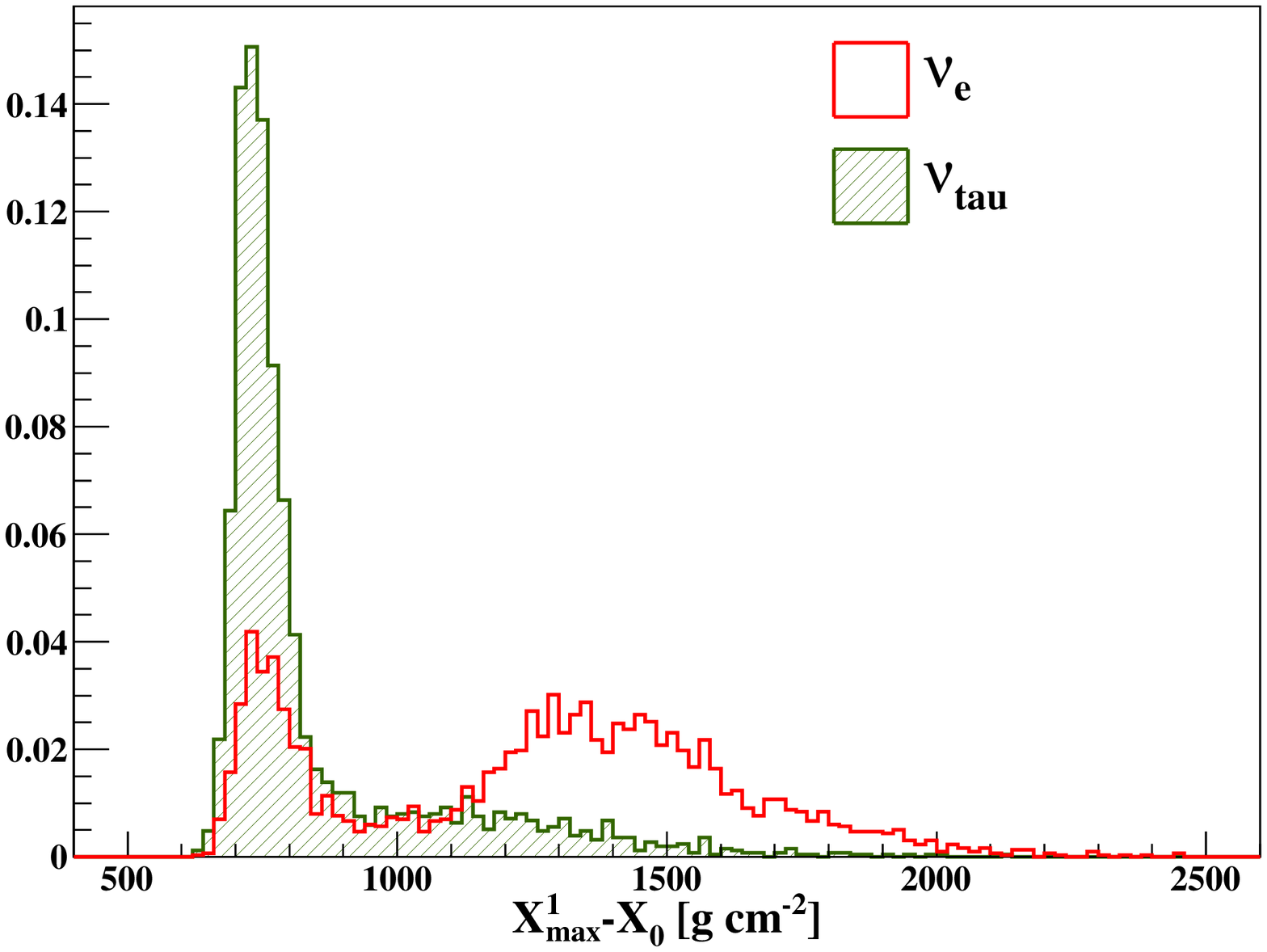}
\includegraphics[width=8cm]{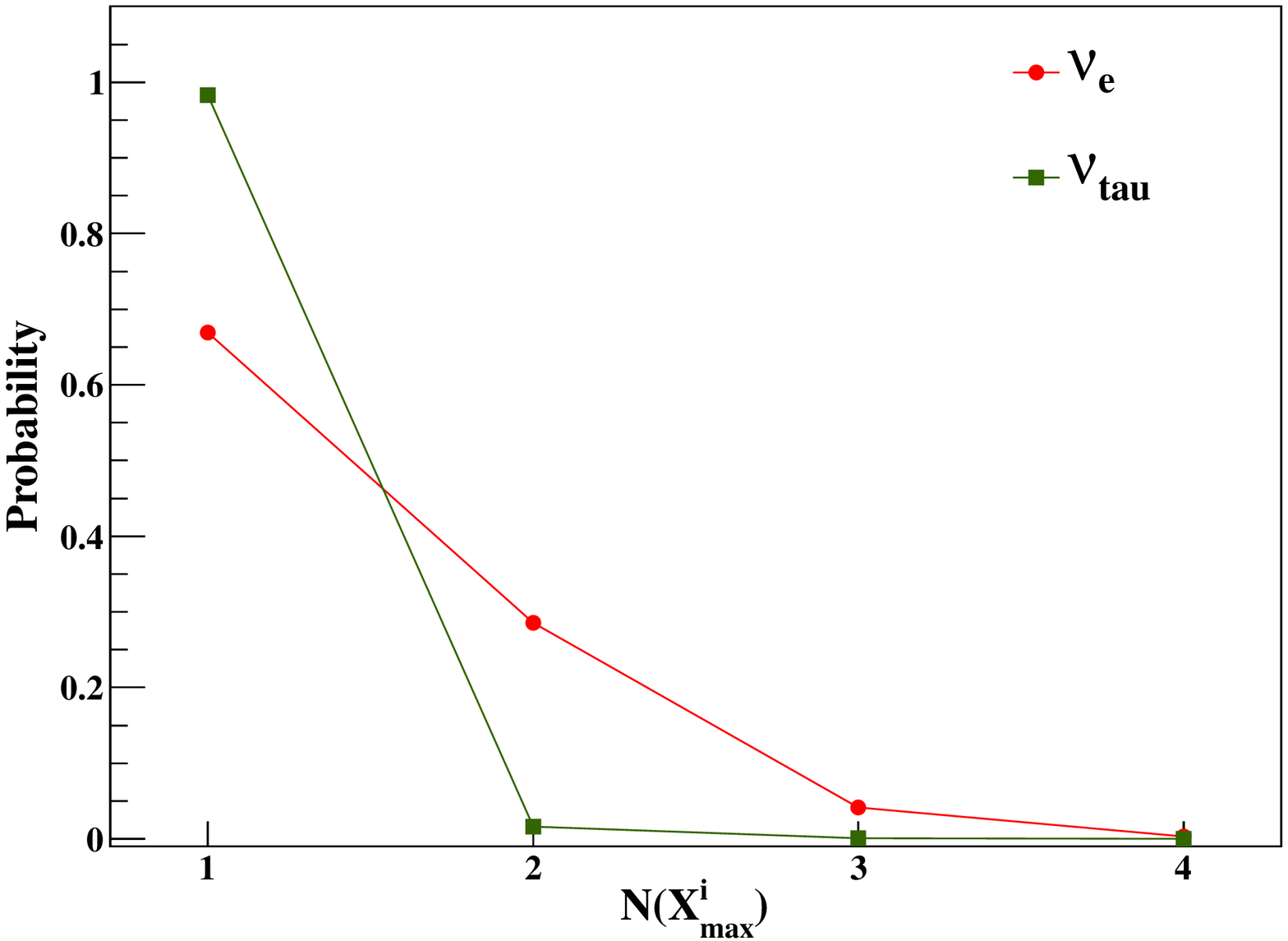}
\caption{\label{XmaxI} Upper panel: distribution of the first peak $\Delta X_{max}^1$ for VDH electron and tau neutrino  
showers. Bottom panel: probability to find $N(X_{max}^i)$ peaks in a profile corresponding to a VDH electron and tau 
neutrino showers. The primary energy of the neutrinos is $E_\nu=10^{20}$ eV.}
\end{figure}

Given a simulated longitudinal profile, the number of peaks and their positions are determined following the procedure 
described in Ref. \cite{Supanitsky:11}. The upper panel of figure \ref{XmaxI} shows the distribution of the position
of the first peak, $X_{max}^1$, found from the beginning of the shower, $X_{0}$, for electron and tau neutrino showers
of $E_\nu=10^{20}$ eV. Note that the distribution function of $\Delta X_{max}^1=X_{max}^1-X_{0}$ for VDH electron 
neutrino showers is bivalued and its first peak is located at $\Delta X_{max}^1 \sim 800$ g cm$^{-2}$, while the 
second one is centered at $\Delta X_{max}^1 \sim 1500$ g cm$^{-2}$. The first peak is related to the development of 
the hadronic component of the cascade, whereas the second one mainly reflects the electromagnetic components of the 
shower (see Ref. \cite{Supanitsky:11} for a detailed discussion). As expected, the distribution function of 
$\Delta X_{max}^1$ for VDH tau neutrino showers is also bivalued but the hadronic peak is much more important. The 
bottom panel of figure \ref{XmaxI} shows the probability to find a profile with $N(X_{max}^i)$ peaks. As expected, 
the profiles corresponding to electron neutrino showers have a larger probability to have more than one peak. In 
particular, the probability to have just one peak for VDH electron neutrino showers of $E_\nu=10^{20}$ eV is $\sim 0.67$, 
whereas for VDH tau neutrino showers it is $\sim 0.98$.  

\begin{figure}[!h]
\includegraphics[width=8cm]{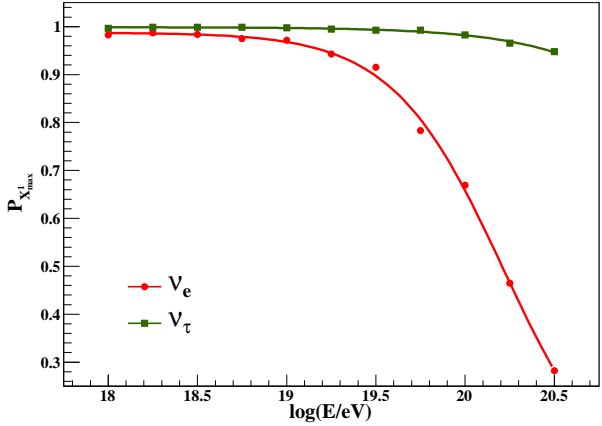}
\caption{Probability to find a profile with just one peak as a function of primary energy for VDH electron and tau neutrino 
showers. \label{PXmaxI}}
\end{figure}

Figure \ref{PXmaxI} shows the probability to find just one peak, $P_{X_{max}^1}$, in a profile corresponding to VDH 
electron and tau neutrino showers as a function of the primary energy. The solid lines in the figure are fits to the 
data points with the function
\begin{equation}
\label{PXmaxIFit}
P_{X_{max}^1}(E_\nu) = \frac{p_0}{1+\left( \frac{E_\nu}{E_0} \right)^{\gamma}},
\end{equation}
where $p_0, E_0$, and $\gamma$ are free fit parameters. The probability to find a profile with just one peak decreases 
with primary energy, owing to the fact that the LPM effect becomes more important at higher energies. As expected, 
$P_{X_{max}^1}(E_\nu)$ for tau neutrinos is larger than the corresponding one for electron neutrinos.

The decay length of a tau lepton of energy $E_\tau$ is given by
\begin{equation}
\label{taudecay}
L_\tau(E_\tau) \cong 4900 \left(  \frac{E_\tau}{10^{20} \textrm{eV}} \right) \textrm{km}.
\end{equation}
Therefore, the probability that a high energy tau, generated in a CC interaction of a horizontal tau neutrino with an 
atmospheric nucleon, decays just above the Earth's surface and is smaller than the probability that an horizontal electron 
neutrino suffers a CC interaction with an atmospheric nucleon at the same point. Because of this, the number of VDH tau 
neutrino showers are suppressed compared with the electron neutrino ones at the higher energies. In particular, the ratio 
between the number of VDH tau neutrino showers and VDH electron neutrino showers for an incident flux with the same number
of electron and tau neutrinos is given by
\begin{eqnarray}
\label{ratiosh}
R_\nu(E_\nu) &=& \frac{dP_{\nu_\tau}^{sh}}{d \xi} (E_\nu,\xi=0) \bigg / \frac{dP_{\nu_e}^{sh}}{d \xi} (E_\nu,\xi=0)  \nonumber \\
&=& \frac{\exp(X_0/\lambda_\nu(E_\nu))}{\rho_0}%
\int_0^\infty dE_\tau\ P(E_\tau| E_\nu)   \nonumber \\
&& \times \int_0^\infty d\xi \exp(-X(\xi)/\lambda_\nu(E_\nu))\ \rho(h(\xi)) \nonumber \\
&& \times \frac{\exp(-\xi/L_\tau(E_\tau))}{L_\tau(E_\tau)}.
\end{eqnarray}
Here $\xi$ is the distance in the horizontal direction measured from a point just above the Earth's surface, 
$dP_{\nu_l}^{sh}/d \xi (E_\nu,\xi=0)$ is the probability per unit of length to initiate an horizontal $l$-flavor neutrino 
shower at $\xi=0$, $P(E_\tau| E_\nu)$ is the probability to generate a tau lepton of energy $E_\tau$ given the CC interaction 
of a tau neutrino of energy $E_\nu$ with a nucleon of the atmosphere, 
$\lambda_\nu(E_\nu)=7.23 \times 10^7 (E_\nu/10^{19}\textrm{eV})^{-0.358}$ g cm$^{-2}$ is the mean free path for CC 
neutrino-nucleon interaction calculated using the cross section of Ref. \cite{Anchordoqui:06}, $\rho(h)$ is the atmospheric 
density as a function of the altitude $h$, 
\begin{equation}
\label{hxi}
h(\xi)=\sqrt{R_\oplus^2+\xi^2}-R_\oplus,
\end{equation}
is the altitude as a function of $\xi$ with $R_\oplus$ the radius of the Earth,
\begin{equation}
\label{Xxi}
X(\xi)=\int_\xi^\infty d\xi' \rho(h(\xi')),
\end{equation}
and $X_0=X(\xi=0)$.

\begin{figure}[!h]
\includegraphics[width=8cm]{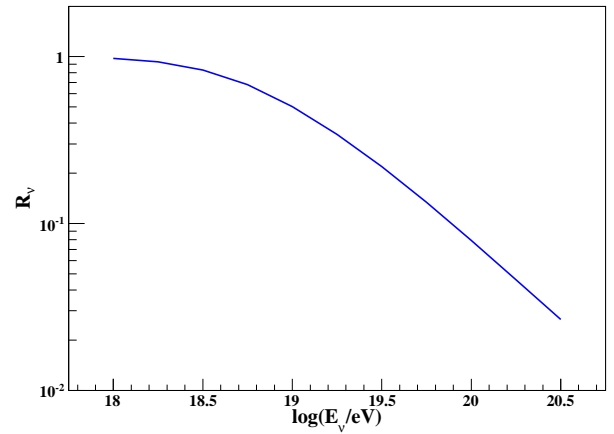}
\caption{\label{Rnu} Ratio of the number of VDH tau neutrino showers to the number of VDH electron neutrino ones as a 
function of primary energy.}
\end{figure}

Figure \ref{Rnu} shows $R_\nu$ as a function of the neutrino energy. The atmospheric density profile used in the calculation 
corresponds to Linsley's parametrization of the US standard model \cite{Aires}, and the probability $P(E_\tau| E_\nu)$ is 
obtained from PYTHIA simulations. $R_\nu$ decreases with the neutrino energy in such a way that between $E_\nu = 10^{18}$ eV 
and $E_\nu = 10^{19.75}$ eV it falls from $\sim 1$ to $\sim 0.13$. Note that when the tau neutrino energy decreases, the energy
of the tau lepton produced in the interaction also decreases; as a consequence the lifetime of the tau lepton, in the laboratory 
frame, also decreases. Therefore, when the tau neutrino energy is small enough, the tau lepton decays almost immediately after being 
produced, and then the shower starts at $\xi \cong 0$ as for the case of electron neutrinos, producing values of $R_\nu$ close 
to one. In any case, for $L_\tau \rightarrow  0$ the function $\exp(-\xi/L_\tau)/L_\tau$, in Eq. (\ref{ratiosh}), can be approximated
by the Dirac delta function, $\delta(\xi)$. Then it can easily be seen that $R_\nu = 1$ when the Dirac function is used in 
Eq. (\ref{ratiosh}).            

\begin{figure}[!h]
\includegraphics[width=8cm]{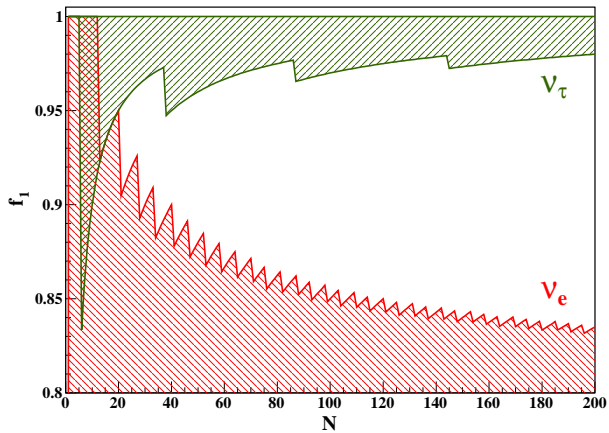}
\caption{\label{Nmin} Regions of $95\%$ probability of the random variable $f_1=n_1/N$ (see the text) as a function the sample 
size $N$ for VDH electron (red) and tau (green) neutrino showers with $E_\nu = 10^{19.75}$ eV.}
\end{figure}

Although the number of VDH tau neutrino showers is smaller compared with the VDH electron neutrino ones, few events can 
give relevant information about the different flavors present in a sample. Figure \ref{Nmin} shows the regions of $95\%$
of probability to find a fraction of neutrino showers with one peak, $f_1=n_1/N$, as a function of the sample size
$N$ for VDH electron and tau neutrino showers of $E_\nu = 10^{19.75}$ eV. For the case of VDH electron neutrino showers 
this region is enclosed by $f_1=0$ and $f_1=f_1^{max}(N)$ and for tau neutrino showers by $f_1=f_1^{min}(N)$ and $f_1=1$ 
where $f_1^{max}(N)$, and $f_1^{min}(N)$ are calculated by using $n_1$ as a binomial random variable. The probability 
to find just one peak in a longitudinal profile is obtained from the fits of $P_{X_{max}^1}$ (see Eq. (\ref{PXmaxIFit})).      
It can be seen that with a sample of a small number of events, it is possible to say something about the neutrino flavors 
present in the sample, depending on the observed number of profiles with just one peak, $n_1$. If $f_1$, obtained from the 
observations, falls in the nonshadowed region, the hypothesis of having a pure sample of electron or tau neutrinos is 
rejected with at least $95\%$ probability. Also for $N\gtrsim 10$, if $f_1$ falls in some of the two shadowed regions, it 
indicates that it is possible to reject the hypothesis of having a pure sample of neutrinos of the opposite flavor, with at 
least $95\%$ probability.

The distribution function of $f_1$ for a given energy, sample size, and the electron neutrino abundance of the incident 
flux for a binary mixture of electron and tau neutrinos, $c_{\nu_e}=N(\nu_e)/(N(\nu_e)+N(\nu_\tau))$, is obtained from
Monte Carlo simulations. Assuming that the distribution function of the number of VDH neutrino showers of a given flavor 
$l$, $n_{\nu_l}$, corresponding to a given observation time is Poissonian, the probability to find $n_{\nu_e}$ and 
$n_{\nu_\tau}$ showers in a sample of size $N = n_{\nu_e}+n_{\nu_\tau}$ is given by the binomial 
distribution
\begin{equation}
\label{Pnush}
P_{sh}(n_{\nu_e}|N) = B(N,p) = \binom{N}{n_{\nu_e}} p^{n_{\nu_e}} (1-p)^{N-n_{\nu_e}},
\end{equation}
where
\begin{equation}
\label{Pnue}
p(E_\nu)=\frac{1}{1+ \mathop{\displaystyle \frac{1-c_{\nu_e}}{c_{\nu_e}}}\ (1-b_\mu)\ R_\nu(E_\nu)}.
\end{equation}
Here the factor $1-b_\mu$ takes into account the fact that the tau neutrino showers that come from the muonic channel of 
the tau decay are not considered. For a given value of the sample size $N$, the number of VDH electron neutrino showers 
is obtained by taking at random a number from the binomial distribution of Eq. (\ref{Pnush}). If in the $i$th step of the 
simulation the number of VDH electron neutrino showers is $n_{\nu_e}^i$ and the number of VDH tau neutrino showers is 
$n_{\nu_\tau}^i=N-n_{\nu_e}^i$, the number of profiles with just one peak is obtained by sampling the binomial distributions 
$B(n_{\nu_e}^i,p_{\nu_e}^1)$  and $B(n_{\nu_\tau}^i,p_{\nu_\tau}^1)$ where $p_{\nu_e}^1$ and $p_{\nu_\tau}^1$ are obtained 
from the fits of $P_{X_{max}^1}$ as explained before.

\begin{figure}[!h]
\includegraphics[width=8cm]{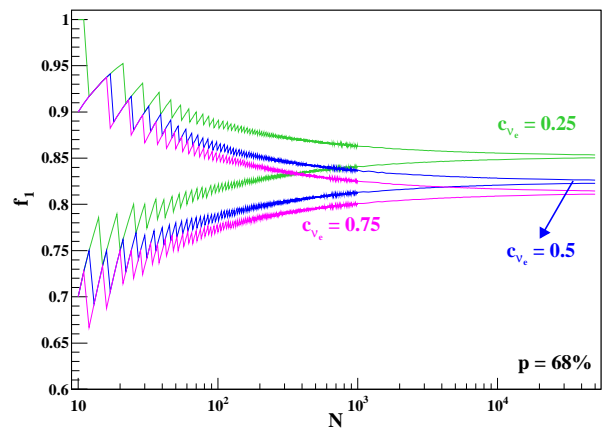}
\caption{\label{Cnue} Regions of $68\%$ probability of $f_1$ for three different values of electron neutrino abundance,
$c_{\nu_e}=\{0.25, 0.5, 0.75\}$. The primary energy corresponds to $E_\nu = 10^{19.75}$ eV.}
\end{figure}

\begin{figure}[!h]
\includegraphics[width=8cm]{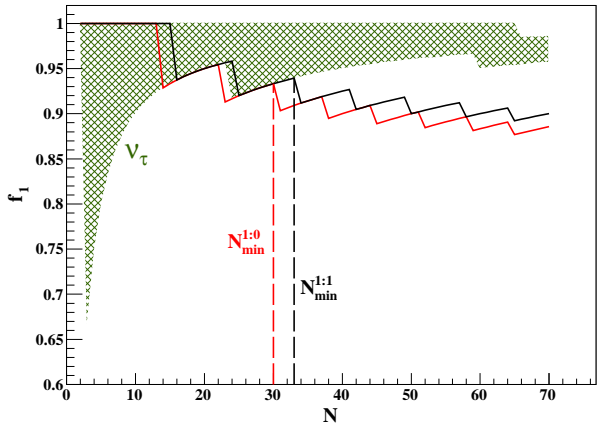}
\caption{\label{f1ex} $f_1$ as a function of $N$ for $E_\nu = 10^{19.75}$ eV. Shadowed region shows the $68\%$ probability
of the distribution function of $f_1$ for the incident flux with flavor content $0:1$. $N_{min}^{1:1}$ and $N_{min}^{1:0}$
correspond to the minimum number of events required to reject the hypothesis $1:1$ and $0:1$ with $95\%$ probability (see 
text for details).}
\end{figure}

Figure \ref{Cnue} shows the intervals of $68\%$ of probability of $f_1$, as a function of the sample size, for three 
different values of electron neutrino abundances, $c_{\nu_e}=\{0.25, 0.5, 0.75\}$, and for $E_\nu = 10^{19.75}$ eV. 
It can be seen that scenarios in which the incident flux is dominated by tau neutrinos are easier to discriminate,
i.e. samples of a smaller size are required. 
Given the flavor content of the incident flux, it is possible to use the parameter $f_1$ to calculate the minimum number 
of events, $N_{min}$, required to reject a false hypothesis. The notation used for the different scenarios considered is
$\Phi_{\nu_e}:\Phi_{\nu_\tau} = 1:1$ for the case in which there are an equal number of electron and tau neutrinos 
on the incident flux, $\Phi_{\nu_e}:\Phi_{\nu_\tau} = 0:1$ means that there are just tau neutrinos, and 
$\Phi_{\nu_e}:\Phi_{\nu_\tau} = 1:0$ means that there are just electron neutrinos. Figure \ref{f1ex} shows an example of 
the calculation of $N_{min}$ for the case in which there are just tau neutrinos on the incident flux and for 
$E_\nu = 10^{19.75}$ eV. The shadowed region corresponds to the region of $68\%$ probability of $f_1$ for the case in which 
there are just tau neutrinos in the sample, i.e., $0:1$. The constant $f_1 = 0$ together with the black line and also $f_1 = 0$ 
together with the red solid line, enclose the regions of $95\%$ probability to find $f_1$ for the cases in which there are an 
equal number of tau and electron neutrinos, $1:1$, and just electron neutrinos, $1:0$, on the incident flux, respectively. Therefore, 
in the $68\%$ of the cases it is possible to reject the hypothesis $1:0$ with $95\%$ probability with a sample of size 
$N>N_{min}^{1:0}=31$ and also, in the $68\%$ of the cases, it is possible to reject the hypothesis $1:1$ with $95\%$ probability 
by using a sample of size $N>N_{min}^{1:1}=33$. Note that, if the region of $95\%$ probability of $f_1$, corresponding to $0:1$, 
is used instead of the region of $68\%$ probability, larger values of $N_{min}$ are obtained.

\begin{figure*}[!ht]
\includegraphics[width=8cm]{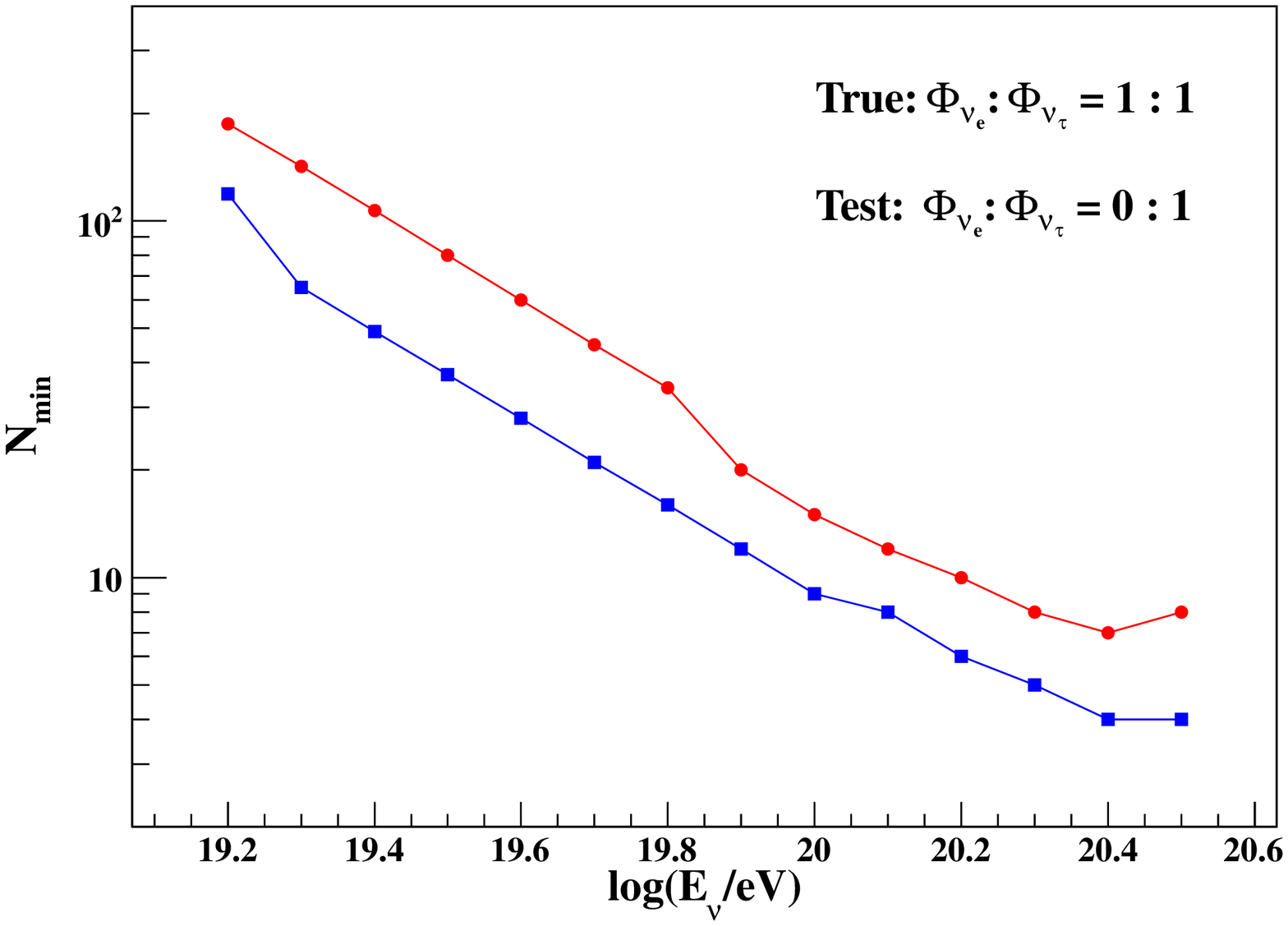}
\includegraphics[width=8cm]{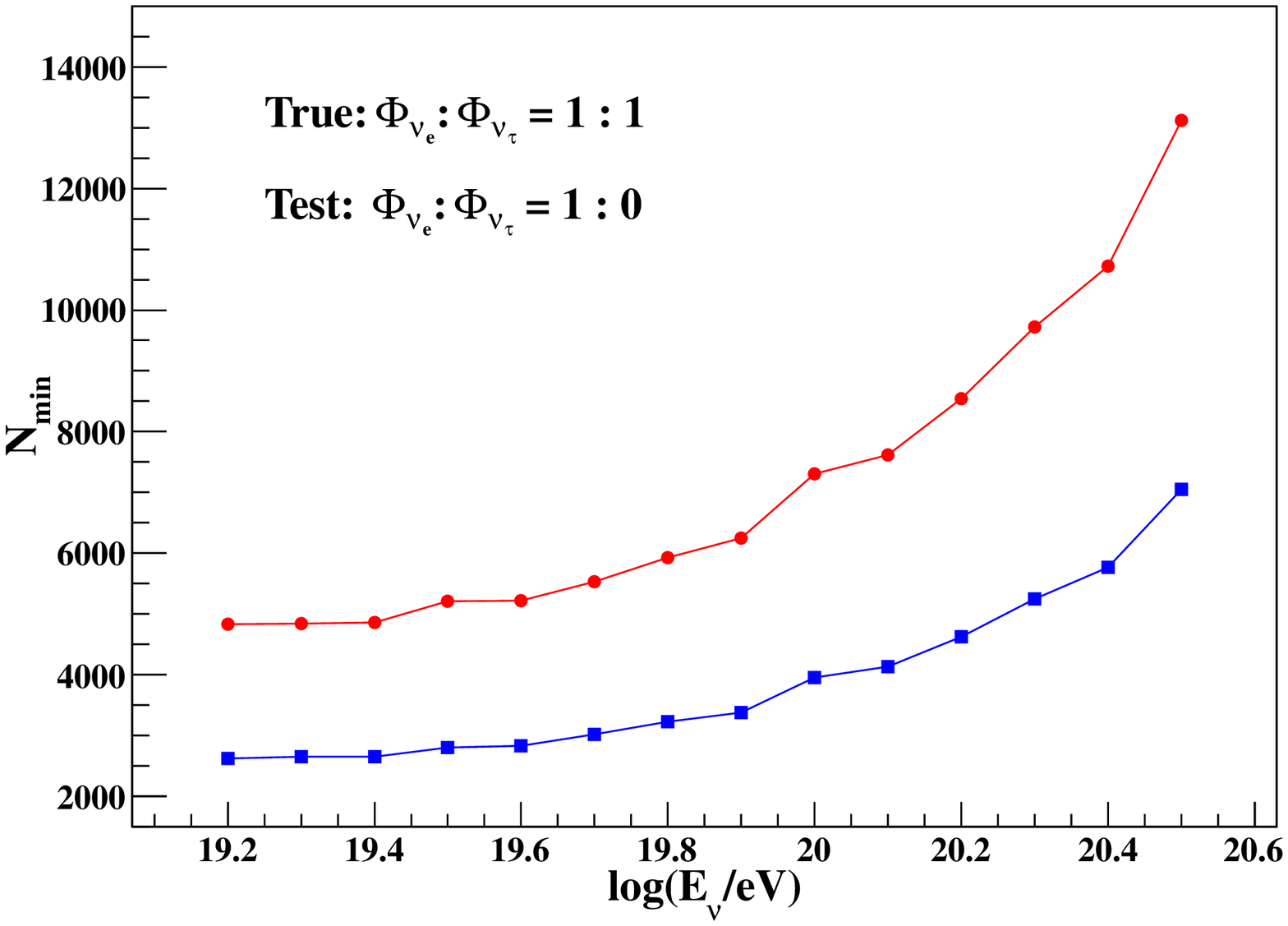}
\includegraphics[width=8cm]{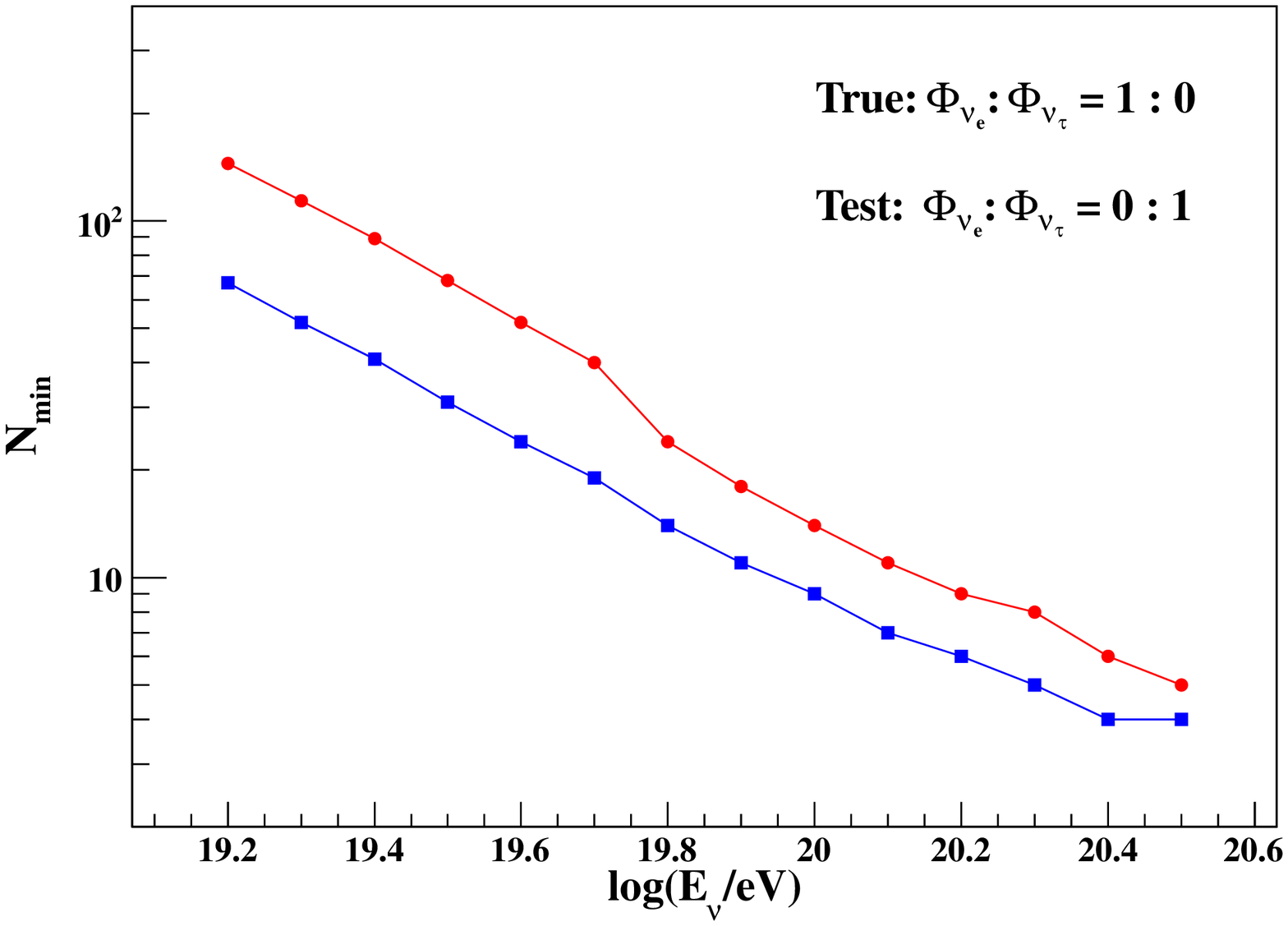}
\includegraphics[width=8cm]{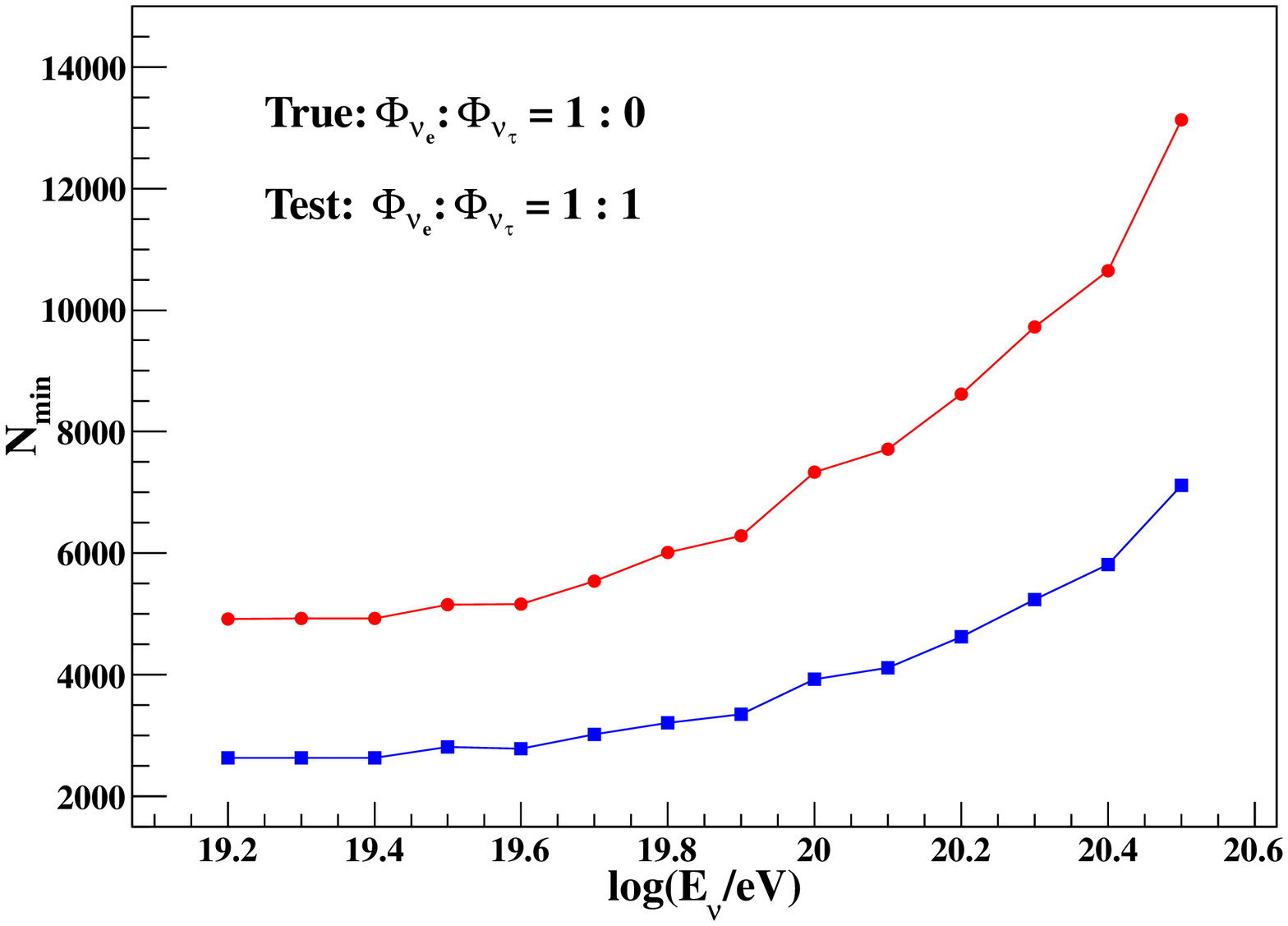}
\includegraphics[width=8cm]{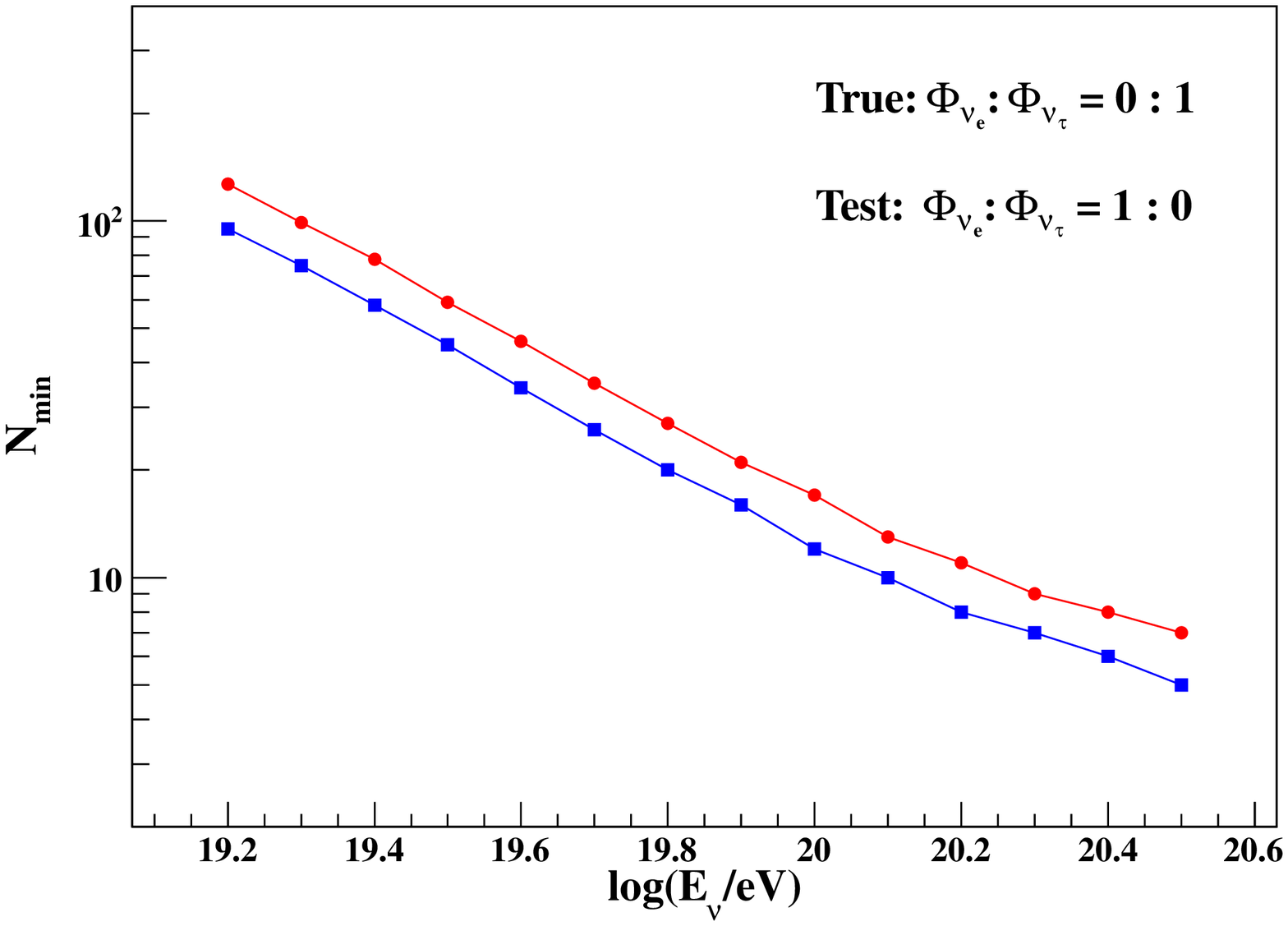}
\includegraphics[width=8cm]{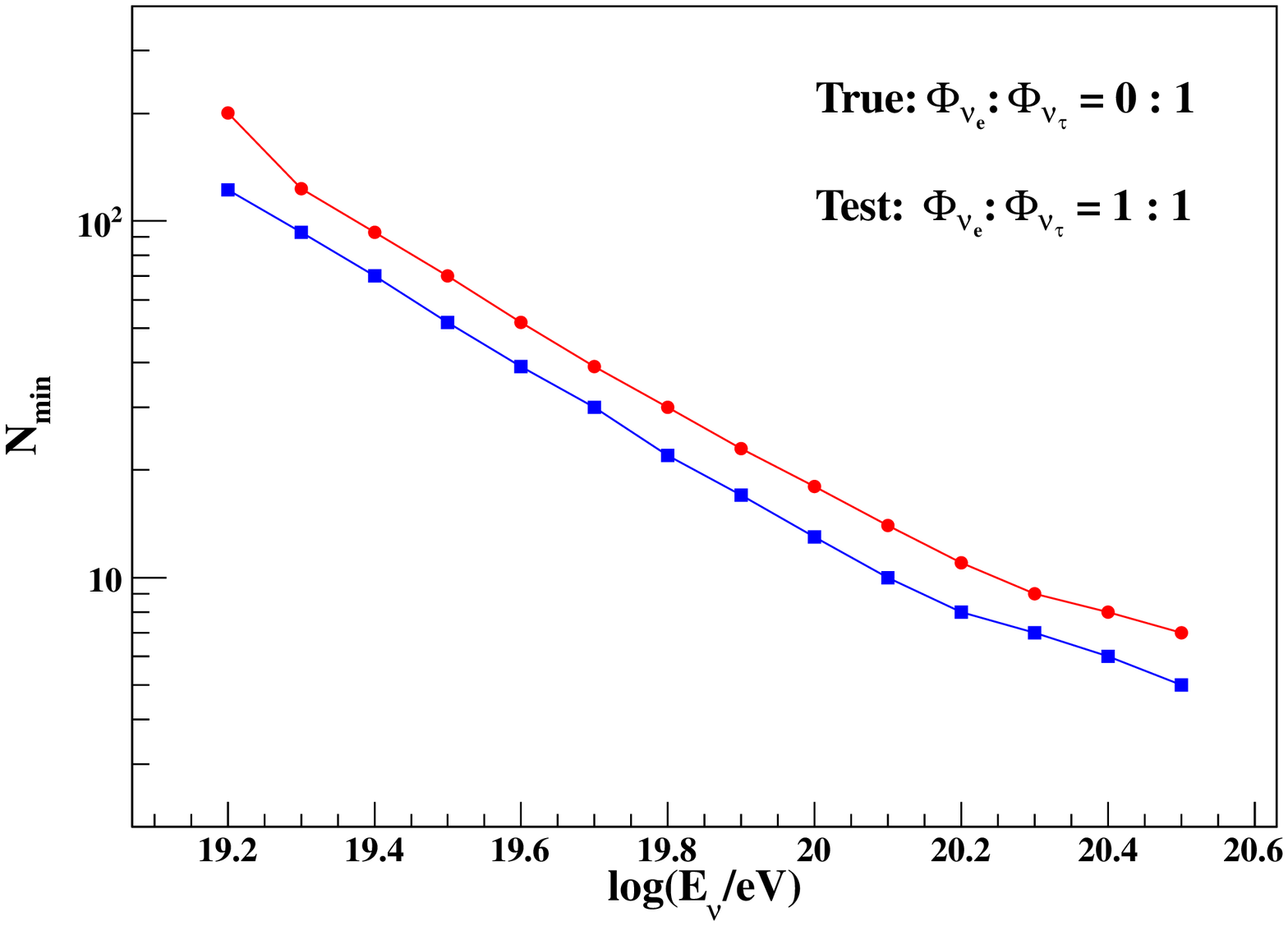}
\caption{\label{NminTests} Minimum number of events, $N_{min}$, as a function of primary energy, needed to reject different 
hypotheses given true scenarios. The red circles correspond to the region of $95\%$ probability of $f_1$ for the true hypothesis
and the blue squares to the region of $68\%$ probability. The rejection probability considered is $95\%$.}
\end{figure*}

Figure \ref{NminTests} shows $N_{min}$ as a function of primary energy for three different hypotheses of the flavor content 
of the incident flux. The rejection probability is $95\%$ and the calculation is done considering the regions of $68\%$ 
probability (blue squares) and $95\%$ probability (red circles) of the distribution function of $f_1$ for the true hypothesis.
$N_{min}$ depends on the difference between the probabilities to find a profile with just one peak corresponding to VDH electron 
and tau neutrino showers, $\Delta p_1 = p_{\nu_\tau}^1-p_{\nu_e}^1$, and also on the parameter $R_\nu$. The larger the values 
of $\Delta p_1$ and the closer to one the values of $R_\nu$ the easier to discriminate among different scenarios. From figures 
\ref{PXmaxI} and \ref{Rnu} it can be seen that while $\Delta p_1$ increases with the neutrino energy, $R_\nu$ decreases because 
of the increase of the tau decay length with primary energy. As a consequence, samples with larger size are required. 

The top panels of figure \ref{NminTests} show the case in which the true hypothesis is $1:1$. It can be seen that the values 
of $N_{min}$ required to reject the hypothesis of having tau neutrinos alone decreases with primary energy. This happens 
mainly because $\Delta p_1$ increases with the neutrino energy and also, but less important, because the distribution function 
of $f_1$, for the $1:1$ case, becomes more similar to the $1:0$ case for increasing values of the primary energy, which is 
caused by the decrease of $R_\nu$ with $E_\nu$. To test the hypothesis of having electron neutrinos alone is more difficult 
for increasing values of $E_\nu$, because, as said before, the distribution function of $f_1$ for the $1:1$ case becomes more 
similar to the corresponding one to the case in which there are electron neutrinos alone on the incident flux. This effect is 
more important than the increase of $\Delta p_1$ with $E_\nu$ causing $N_{min}$ to increase with the neutrino energy. Note that 
the values of $N_{min}$ required to reject the hypothesis of having tau neutrinos alone is more than 2 orders of magnitude 
smaller than the ones required to reject the hypothesis of having electron neutrinos alone. In particular, for $E_\nu = 10^{19.7}$ eV 
and considering the regions of $95\%$ probability of $f_1$ for the true hypothesis, $N_{min} \cong 45$ to reject the hypothesis 
$0:1$ with $95\%$ probability and $N_{min} \cong 5529$ to reject the hypothesis $1:0$ with $95\%$ probability.
           
The middle panels of figure \ref{NminTests} show the case in which the incident flux is composed by electron neutrinos only, $1:0$. 
For similar reasons discussed for the previous case, $N_{min}$ is a decreasing function of energy when the hypothesis $0:1$ is 
considered. $N_{min}$ is an increasing function of energy when the hypothesis $1:1$ is considered and $N_{min}$ is more than 
2 order of magnitude larger compared with the corresponding one to the test of hypothesis $0:1$.    

Finally, the bottom panels of figure \ref{NminTests} show the scenario in which the incident flux is composed by tau neutrinos
alone. In both cases $N_{min}$ is a decreasing function of energy because of the increase of $\Delta p_1$ with primary energy.
As expected, to reject the hypothesis $1:1$ requires a larger number of events because of the contribution of tau neutrinos
to $f_1$. In any case, at $E_\nu = 10^{18}$ eV for $95\%$ of the cases, it is possible to reject both hypotheses with 
$\lesssim 200$ events and with rejection probability of 95\% events. At $E_\nu = 10^{20}$ eV just $\lesssim 20$ events are 
required.  

Note that the energy deposition of electron and tau neutrino showers is very different because the tau leptons, generated 
in the CC interactions of tau neutrinos with the nucleons of the atmosphere, decay to tau neutrinos and other particles; 
thus the energy taken by this secondary tau neutrino do not go to the shower. Therefore, if the deposited energy is used 
to reconstruct the primary energy without any correction, the standard procedure when fluorescence telescopes are used to 
observe the showers, then tau showers are reconstructed with smaller energies than that corresponding to electron neutrinos. 
The impact of the energy determination on the method developed above is studied in detail in Appendix.

\section{Conclusions}

In this work we have developed a new statistical technique intended to study the flavor content of the incident flux of
high energy astrophysical neutrinos. This new method is based on the morphological differences between the longitudinal 
profiles corresponding to very deep horizontal electron and tau neutrino showers that can be observed by means of 
fluorescence telescopes. In particular, the multipeak structure of the profiles strongly depends on the flavor of the 
incident neutrino. Very deep horizontal showers initiated by electron neutrinos present more fluctuations, owing to the LPM 
effect, than the corresponding one to tau neutrinos. Therefore, the number of showers with just one peak can be used to 
discriminate between different scenarios for the flavor content of the incident neutrino flux.

The difference between the probability to find just one peak, in a given profile, for very deep horizontal electron and 
tau neutrino showers starts to be important for energies $\gtrsim 10^{19.5}$ eV. Then this technique is more relevant 
for fluorescence telescopes in orbit around the Earth, like the upcoming JEM-EUSO mission and Super-EUSO, owing to their 
huge exposure.      

Scenarios in which the incident flux is dominated by tau neutrinos are easy to identify, and samples with several 
events, depending on the primary energy, are required in order to reject the hypothesis of having electron neutrinos 
alone. However, because of the suppression of very deep horizontal tau neutrino showers with respect to electron neutrino 
ones, a large number of events is required to discriminate between scenarios with electron neutrinos alone and a 
mix of an equal number of tau and electron neutrinos.

\begin{acknowledgments}

A.D.S. is a  members of the Carrera del Investigador Cient\'ifico of CONICET, Argentina. G.M.T. acknowledges the 
support of PAPIIT-UNAM through Grant No. IN-115210-3 and CONACyT through Grant No. CB-2007-83539. This work is part 
of the ongoing effort for the design and development of the JEM-EUSO mission and the definition of its scientific 
objectives.

\end{acknowledgments}

\appendix

\section{Energy deposition of neutrino showers}
\label{appEd}

\begin{figure}[!h]
\includegraphics[width=8cm]{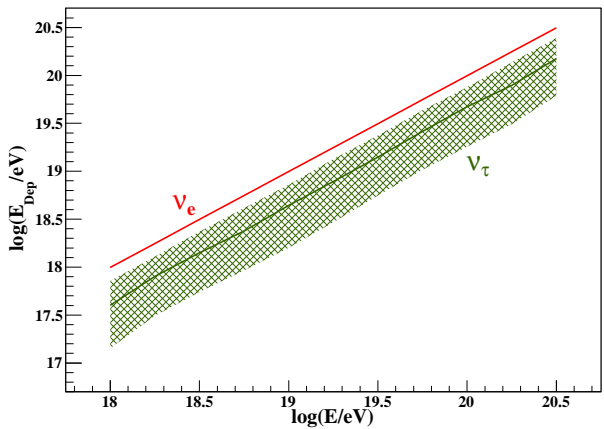}
\caption{\label{Ed} Deposited energy of VDH electron and tau neutrino showers as a function of primary energy. The solid 
lines are the median of the distributions and the shadowed regions correspond to the regions of $68\%$ probability.}
\end{figure}

Tau neutrino showers are initiated by the decay of the tau leptons, generated in CC interactions of tau neutrinos 
with atmospheric nucleons. As mentioned before, the generated tau lepton takes part of the energy of the incident tau 
neutrino. Also, in all decay modes of the taus, a tau neutrino is generated that takes part of the energy of the tau. 
This energy is missed because the neutrinos do not participate in the shower development. Figure \ref{Ed} shows the 
median and the region of $68\%$ probability of the deposited energy as a function of the incident neutrino energy for 
VDH tau and electron neutrino showers. The deposited energy is obtained by integrating the longitudinal profiles of 
the showers. The deposited energy is the observable used to infer the energy of the primary particle when the 
fluorescence telescopes are used to observe the air showers. From the figure it can be seen that energy deposition of 
tau neutrino showers is smaller than the corresponding one to electron neutrino showers with the same primary energy. 
Also the fluctuations are much larger for the case of tau neutrinos.  

\begin{figure}[!h]
\includegraphics[width=8cm]{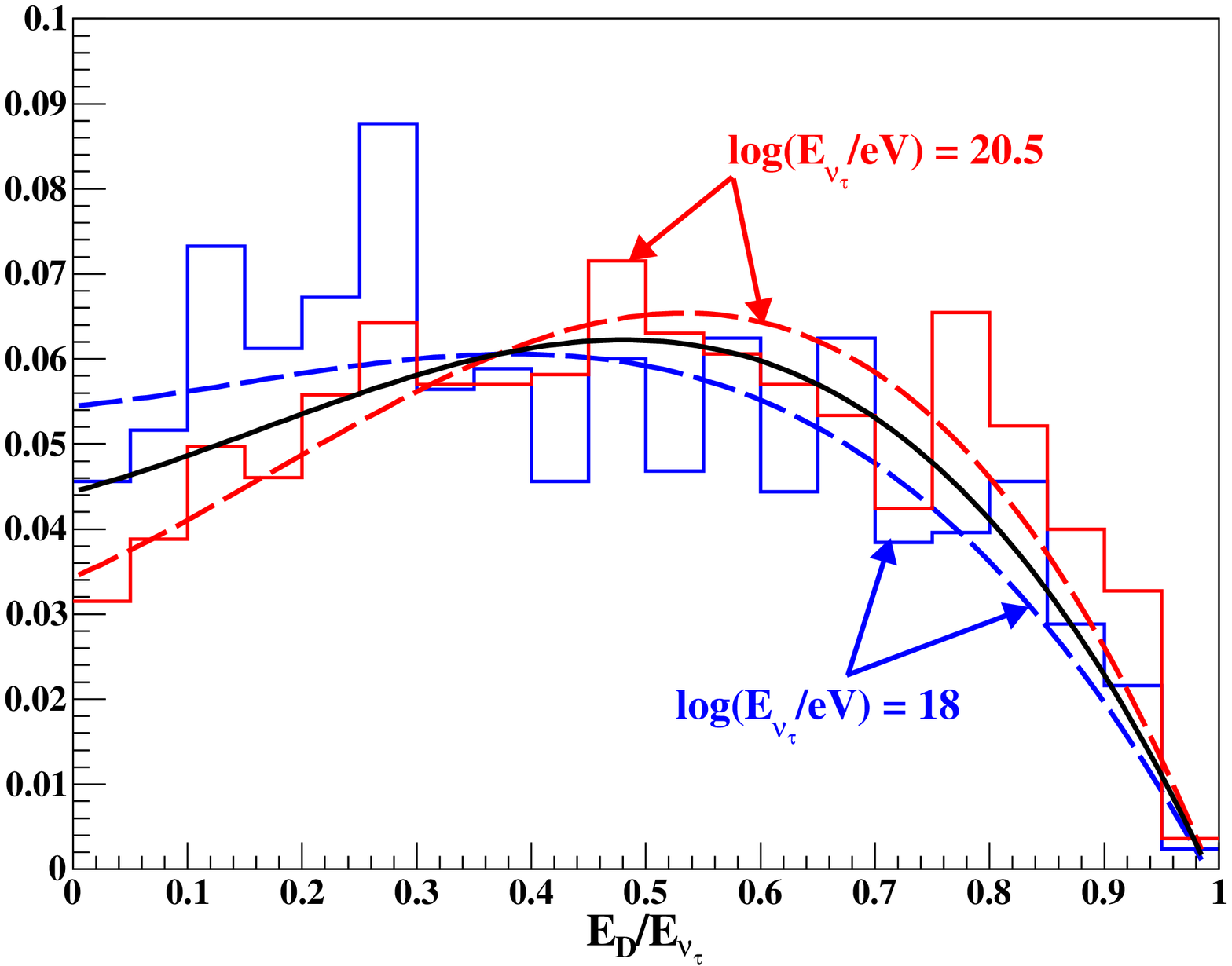}
\includegraphics[width=8cm]{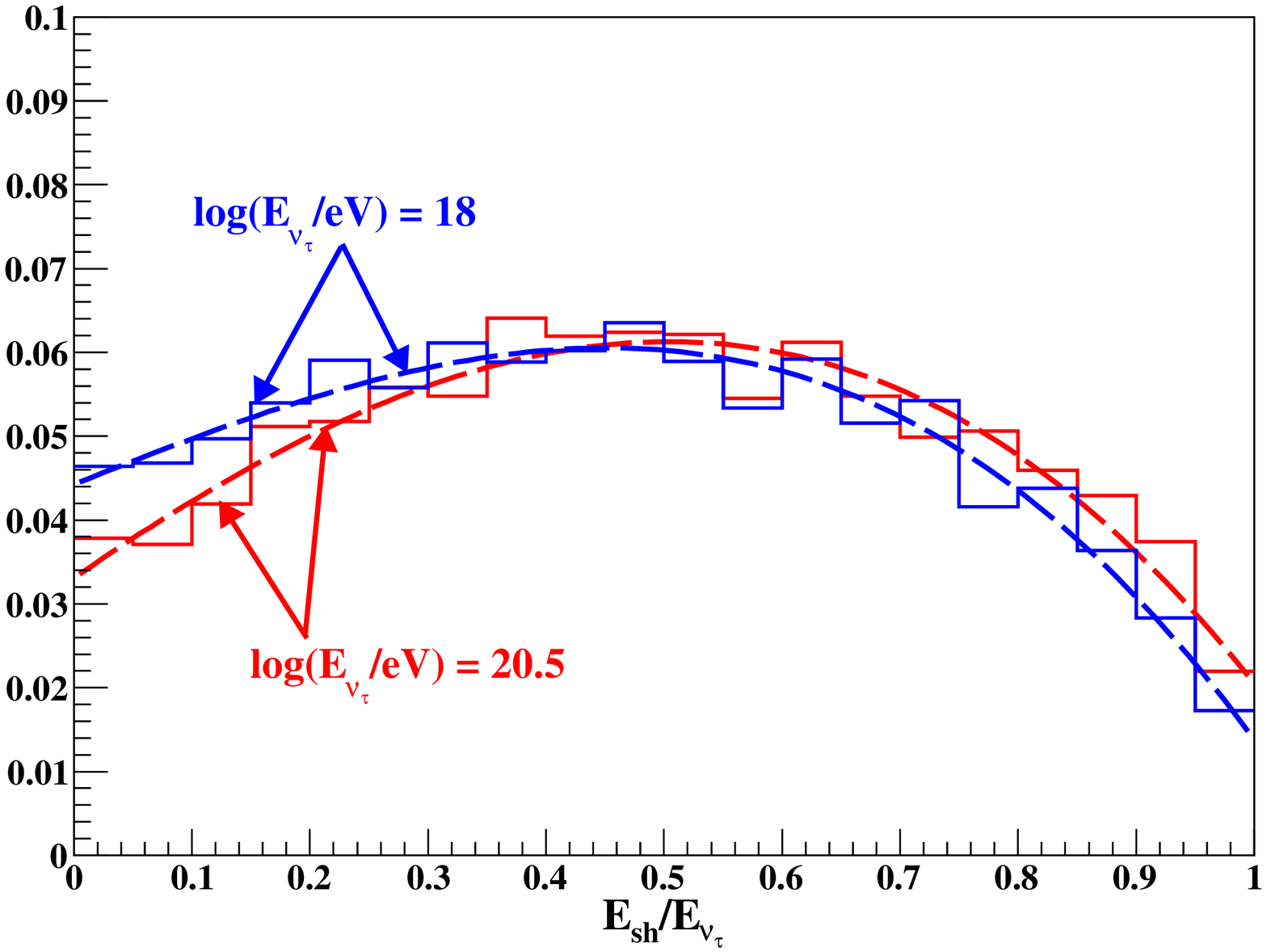}
\caption{\label{FrFit} Top panel: distribution of the ratio between the deposited energy and the tau neutrino energy
for $E_{\nu_\tau} = 10^{18}$ eV and $E_{\nu_\tau} = 10^{20.5}$ eV. Bottom panel: distribution of the ratio between the 
energy that goes to the shower (see text) and the tau neutrino energy for $E_{\nu_\tau} = 10^{18}$ eV and 
$E_{\nu_\tau} = 10^{20.5}$ eV. Dashed lines are fits of the distributions with third degree polynomials, and the solid 
line in the top panel is the average of the two fits.}
\end{figure}

The top panel of figure \ref{FrFit} shows the distribution of the ratio between the deposited energy, $E_{D}$, and the 
tau neutrino energy, $E_{\nu_\tau}$, for $E_{\nu_\tau} = 10^{18}$ eV and $E_{\nu_\tau} = 10^{20.5}$ eV, the minimum and 
the maximum values of energy considered, respectively. It can be seen that the shapes of these distributions do not change 
significantly with the tau neutrino energy. The large fluctuations on the deposited energy of tau neutrino showers are 
dominated by the fluctuations of the energy taken by the tau neutrino generated in the tau decays. The latter can be 
inferred from the plot at the bottom panel of figure \ref{FrFit}, which shows the distribution of the ratio between the 
energy that goes to the shower, $E_{sh}$, and the tau neutrino energy for the same values of $E_{\nu_\tau}$ considered 
before. $E_{sh}$ is obtained by adding the energy of the particles generated in the tau decays except the one of neutrinos. 
It can be seen that the energy that goes to the showers presents large fluctuations and also that, on average, the deposited
energy is smaller than $E_{sh}$, $8\%$ for $E_{\nu_\tau} = 10^{18}$ eV and $2\%$ for $E_{\nu_\tau} = 10^{20.5}$ eV. This is 
due to the fact that there is also missing energy in the development of the shower.   

For the case of electron neutrinos the deposited energy is smaller than the primary energy by less than $\sim 1\%$, and 
the fluctuations, $\sigma[E_D]/\textrm{med}(E_D)$, are smaller than $\sim 0.3\%$ in the whole energy range under 
consideration.   

As mentioned before, the deposited energy of the showers is used to infer the primary energy. Thus, without any correction, 
the reconstructed energy of tau neutrino showers is smaller than the corresponding one to electron neutrino showers. On 
average, the deposited energy of tau neutrino showers goes from $\sim 0.4\ E_{\nu_\tau}$ at $E_{\nu_\tau} = 10^{18}$ eV 
to $\sim 0.5\ E_{\nu_\tau}$ at $E_{\nu_\tau} = 10^{20.5}$ eV. Therefore, because of the fact that $P_{X_{max}^1}$ decreases 
with the neutrino energy, the difference between $P_{X_{max}^1}$ for tau and electron neutrinos becomes smaller when the 
deposited energy is used to estimate the neutrino energy, decreasing the discrimination power of $f_1$.

The distribution of $E_D/E_{\nu_\tau}$ can be fitted by a third degree polynomial, as can be seen from the top panel of 
figure \ref{FrFit}. Because of the small variation of the shape of this distribution with $E_{\nu_\tau}$, the average between
the fits of the distributions corresponding to $E_{\nu_\tau} = 10^{18}$ eV and $E_{\nu_\tau} = 10^{20.5}$ eV is used for 
the whole energy range under consideration. The solid line on the plot of the top panel of figure \ref{FrFit} corresponds 
to the average function, denoted as $P(E_D|E_{\nu_\tau})$. 

The probability to find just one peak in a given profile as a function of the deposited energy is given by,
\begin{equation}
\label{EqP1Ed}
\bar{P}_{X_{max}^1}(E_D) = \int_0^\infty dE_{\nu_\tau} P_{X_{max}^1}(E_{\nu_\tau})\ P(E_{\nu_\tau}|E_D)
\end{equation}   
where 
\begin{equation}
\label{Penued}
P(E_{\nu_\tau}|E_D) = \frac{P(E_D|E_{\nu_\tau})\ P(E_{\nu_\tau})}{\int_0^\infty dE_{\nu_\tau} P(E_D|E_{\nu_\tau})\ P(E_{\nu_\tau})}.
\end{equation}
Here, the energy distribution of the neutrinos is taken as a power law, $P(E_{\nu_\tau})=K E_{\nu_\tau}^{-\gamma}$.

For the case of electron neutrinos, the approximation $P(E_D|E_{\nu_e}) \cong \delta (E_D-0.99\ E_{\nu_e})$ is used. Here 
$\delta(x)$ is the Dirac delta function. By using Eq. (\ref{EqP1Ed}) it is easy to show that 
$\bar{P}_{X_{max}^1}(E_D) \cong P_{X_{max}^1}(E_D/0.99)$. 

\begin{figure}[!ht]
\includegraphics[width=8cm]{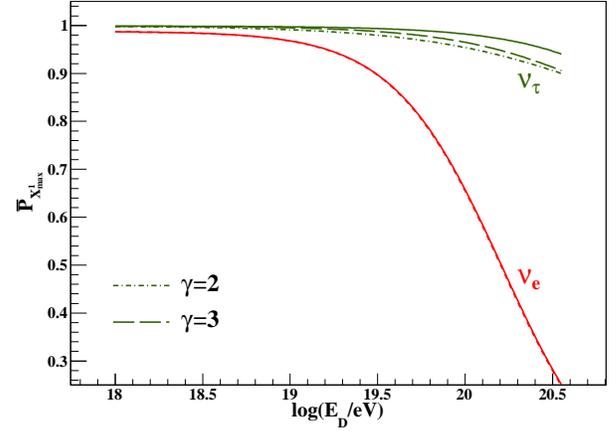}
\caption{\label{P1Ed} $\bar{P}_{X_{max}^1}$ as a function of $\log(E_D/\textrm{eV})$. The red lines correspond to VDH electron 
neutrino showers and the green lines to VDH tau neutrino showers. The solid lines correspond to the ideal case in which 
$E_D = E_{\nu}$. The dashed red line corresponds to VDH electron neutrino showers for which the approximation $E_D = 0.99\ E_{\nu}$ 
is used.}
\end{figure}

Figure \ref{P1Ed} shows $\bar{P}_{X_{max}^1}(E_D)$ for VDH electron and tau neutrino showers as a function of the deposited energy. 
The solid lines correspond to the ideal case in which $E_D = E_{\nu}$; i.e., they correspond to the fits shown in figure \ref{PXmaxI}. 
It can be seen that, for VDH electron neutrino showers, $\bar{P}_{X_{max}^1}(E_D)$ is practically indistinguishable from 
$P_{X_{max}^1}(E_D)$. However, for the case of VDH tau neutrino showers $\bar{P}_{X_{max}^1}(E_D)$ is smaller than the ideal case 
and the difference increases with the deposited energy. Note that, for the case of VDH tau neutrino showers, $\bar{P}_{X_{max}^1}(E_D)$ 
depends on the energy distribution of the incident flux (see Eq. (\ref{Penued})). From the figure it can be seen that, as expected, the 
steeper the neutrino flux the smaller the difference with the ideal case.   

The ratio between the number of VDH tau neutrino showers and the number of VDH electron neutrino ones, as a function of the deposited 
energy, is given by (see Eq. (\ref{ratiosh}))
\begin{equation}
\label{ratioshed}
\bar{R}_\nu(E_\nu) = \frac{d\bar{P}_{\nu_\tau}^{sh}}{d \xi} (E_D,\xi=0) \bigg / \frac{d\bar{P}_{\nu_e}^{sh}}{d \xi} (E_D,\xi=0)  
\end{equation}
where
\begin{eqnarray}
\label{Pshed}
\frac{d\bar{P}_{\nu_\tau}^{sh}}{d \xi} (E_D,0) &=&  \! \! \int_0^\infty  \! \! \! \! \! \! dE_{\nu_\tau} \frac{dP_{\nu_\tau}^{sh}}{d \xi} (E_{\nu_\tau},0)%
P(E_{\nu_\tau}|E_D) \! , \\ 
\frac{d\bar{P}_{\nu_e}^{sh}}{d \xi} (E_D,0)  &=& \frac{dP_{\nu_e}^{sh}}{d \xi} (E_D/0.99,0). 
\end{eqnarray}

\begin{figure}[t]
\includegraphics[width=8cm]{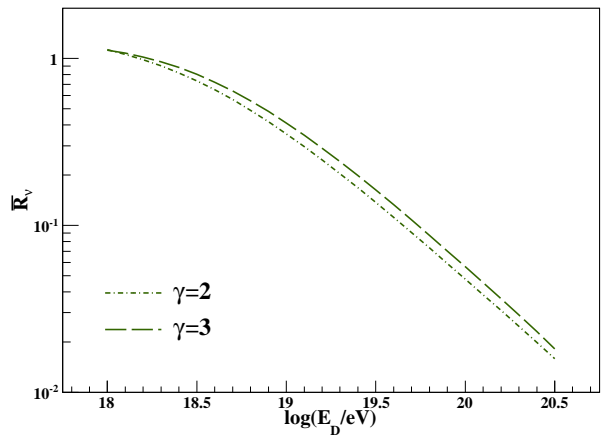}
\caption{\label{RnuEd} Ratio between the number of VDH tau neutrino showers and the number of VDH electron neutrino ones as a 
function of the deposited energy.}
\end{figure}

Figure \ref{RnuEd} shows $\bar{R}_\nu(E_D)$ as a function of the deposited energy for $\gamma = 2$ and $\gamma = 3$. For
$\log(E_D/\textrm{eV}) \gtrsim 18.5$, $\bar{R}_\nu(E)$ is smaller than $R_\nu(E)$.

\begin{figure*}[!]
\includegraphics[width=8cm]{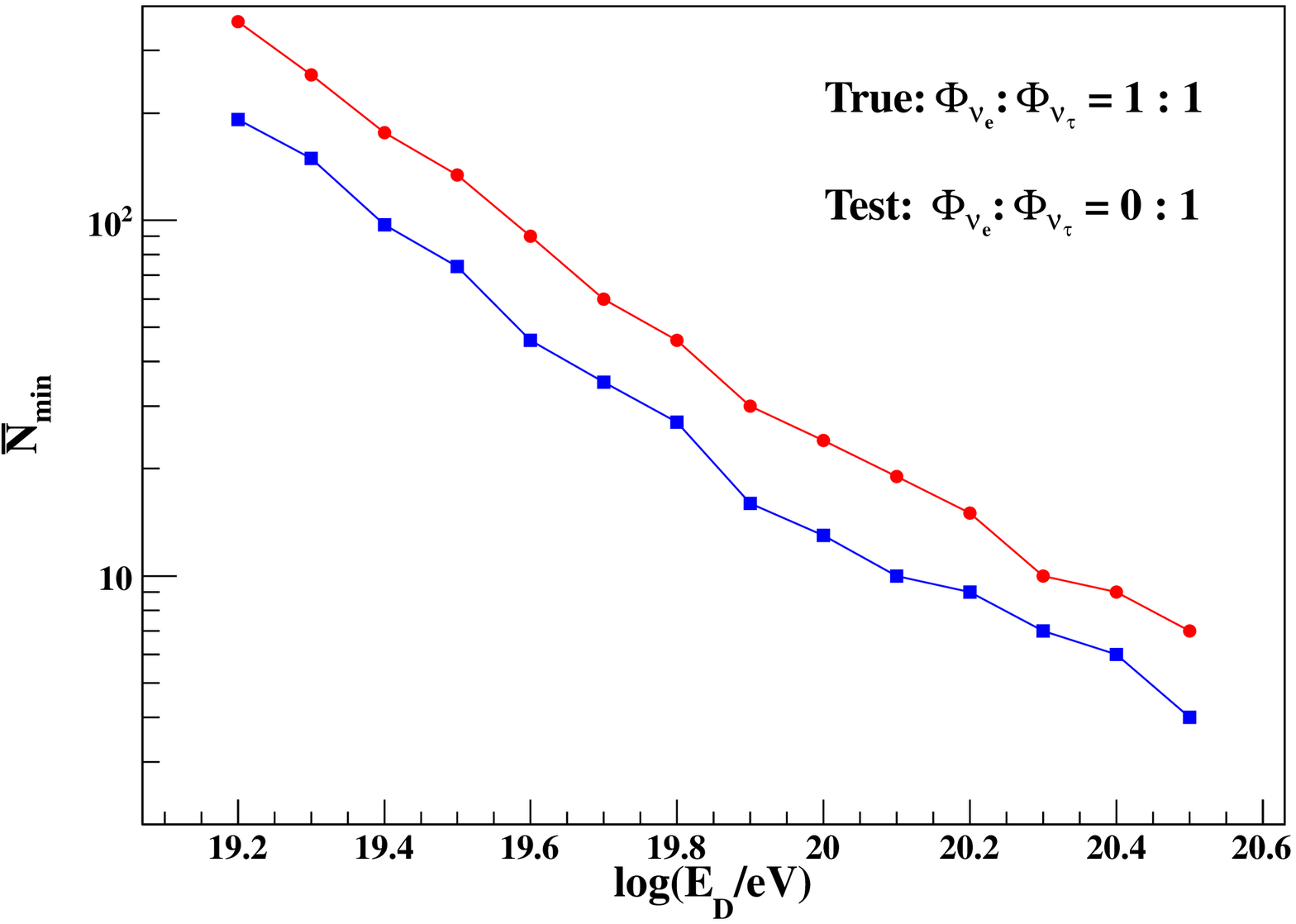}
\includegraphics[width=8cm]{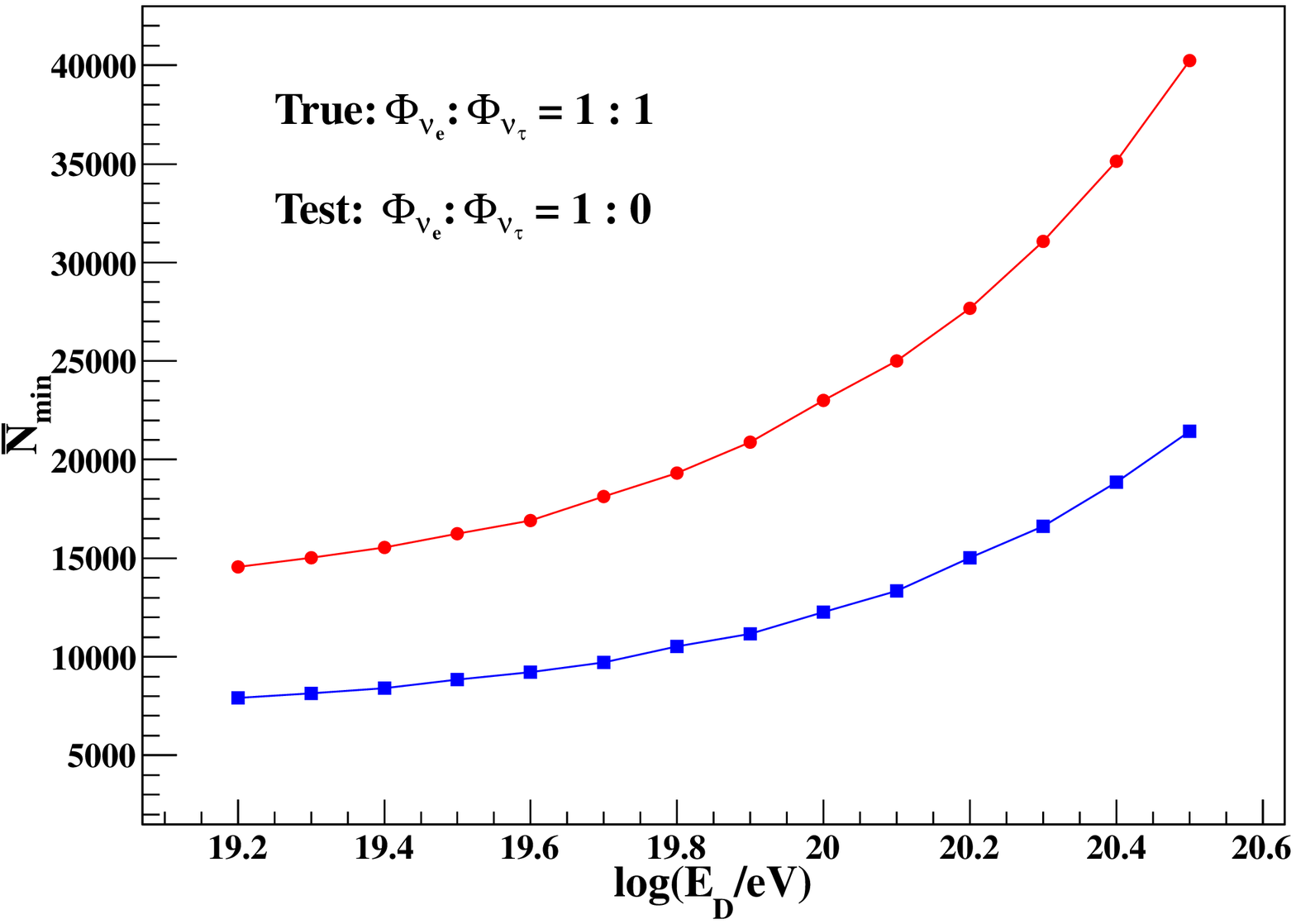}
\includegraphics[width=8cm]{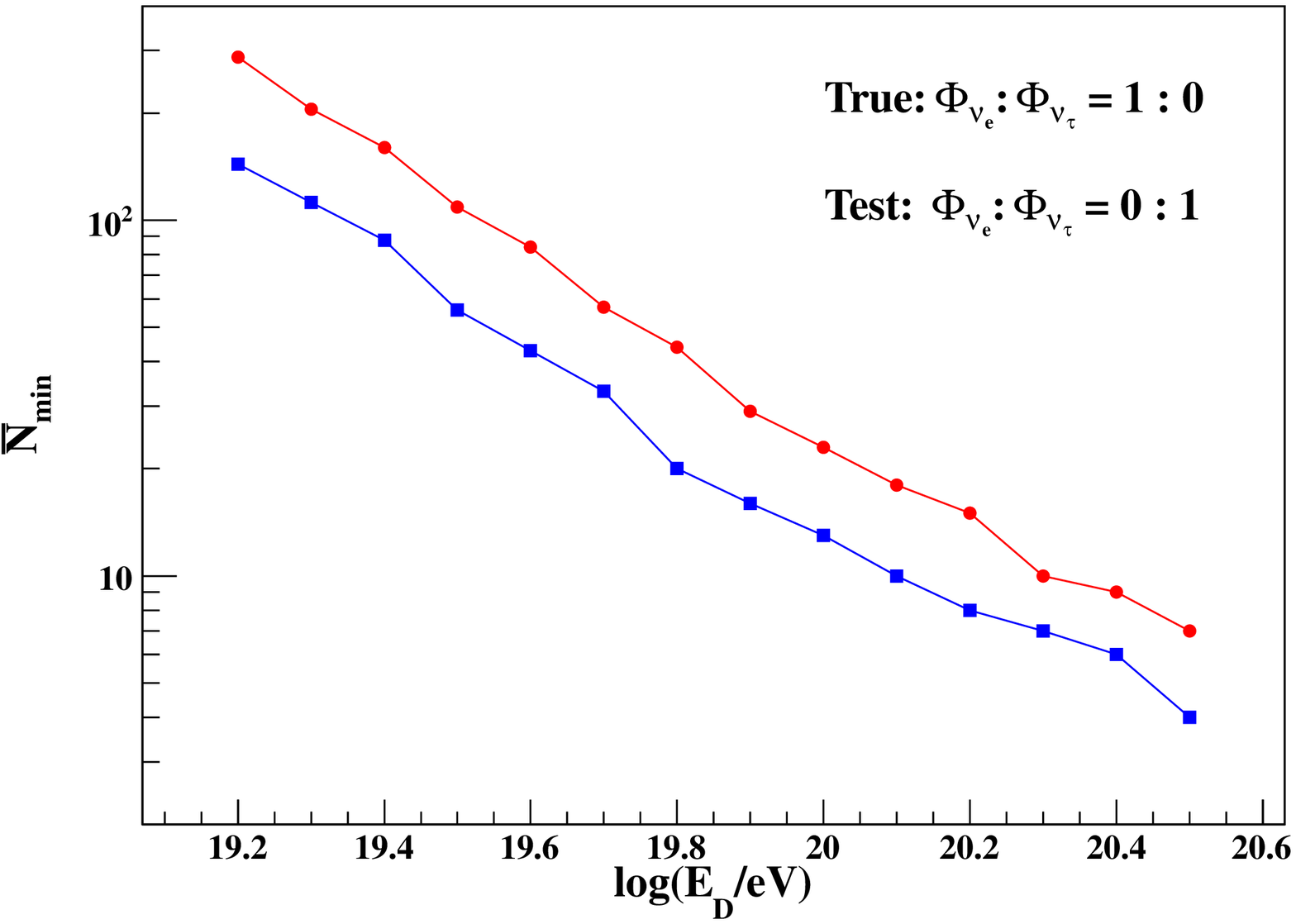}
\includegraphics[width=8cm]{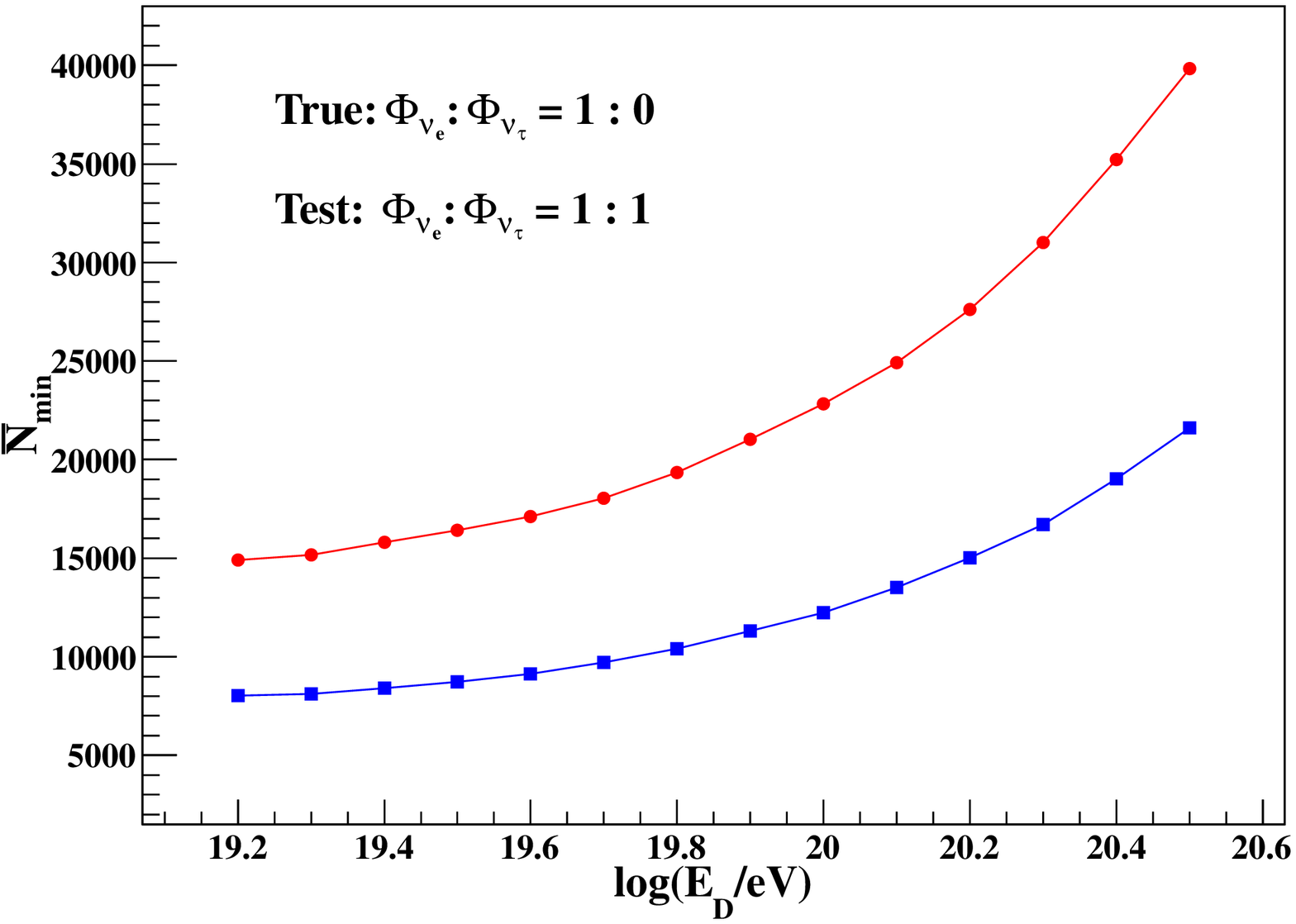}
\includegraphics[width=8cm]{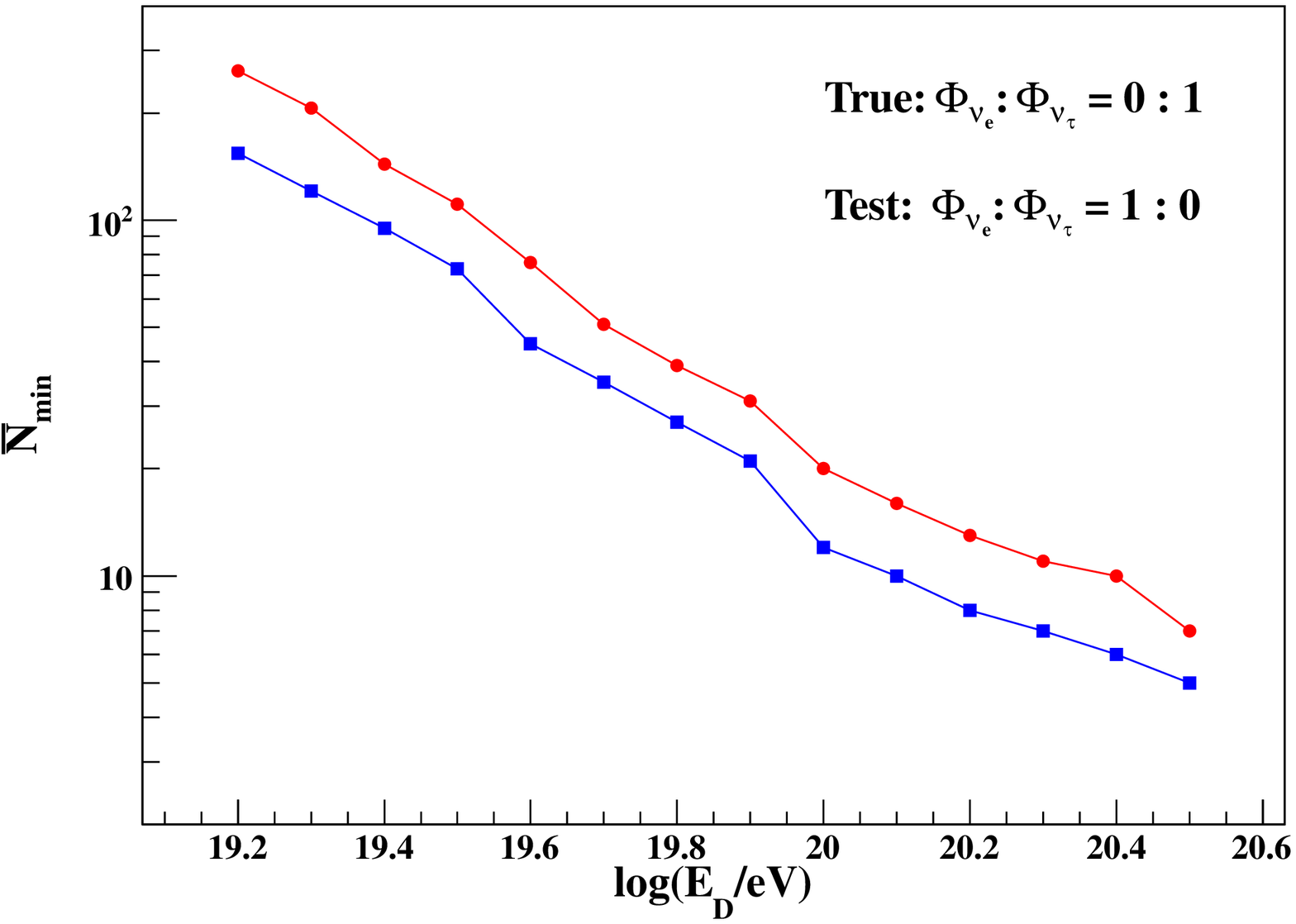}
\includegraphics[width=8cm]{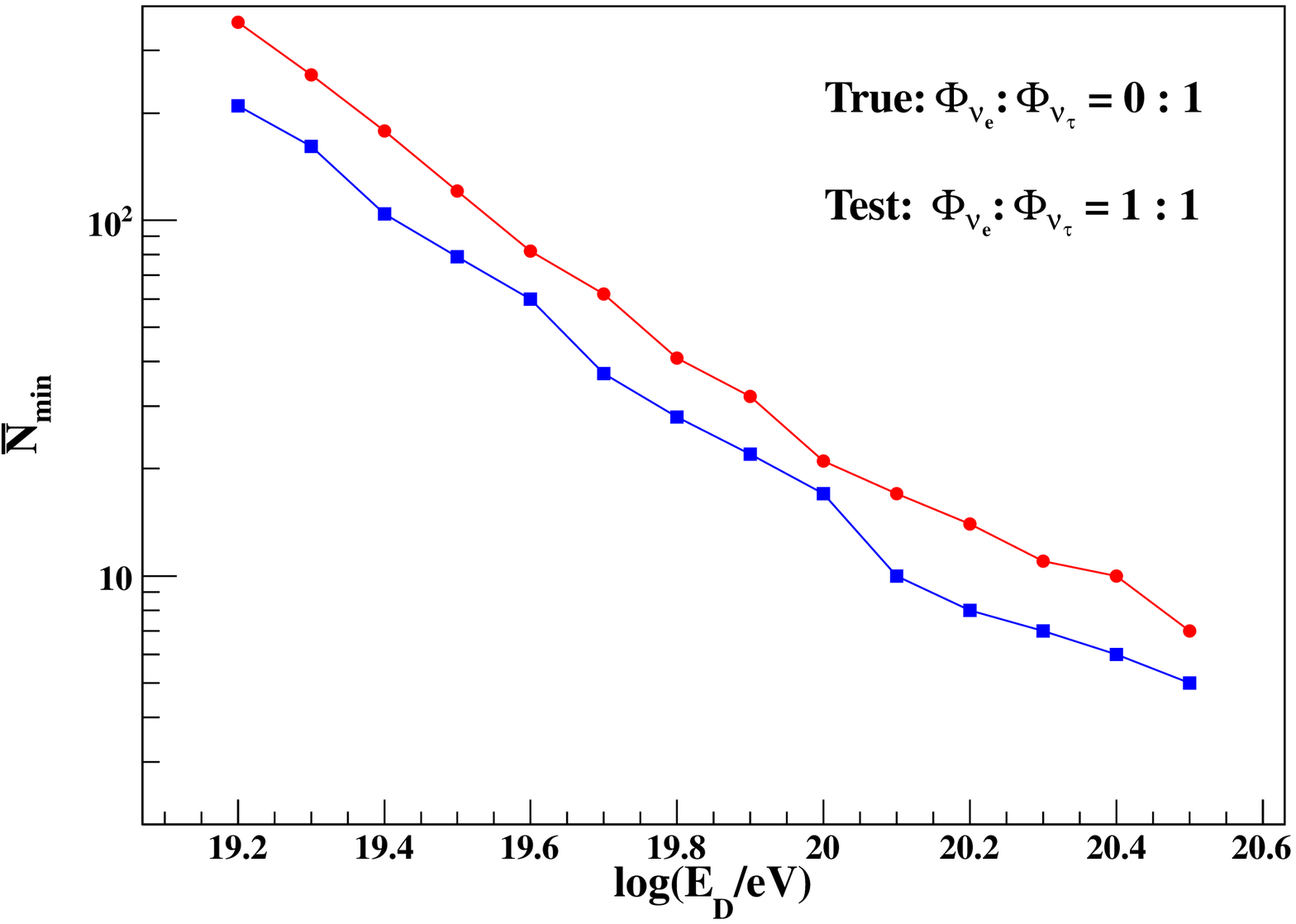}
\caption{\label{NminTestsEd} Minimum number of events, $\bar{N}_{min}$, as a function of the deposited energy, needed to reject 
different hypotheses given true scenarios. The energy distribution of the incident neutrino flux corresponds to a power law of 
spectral index $\gamma=2$.}
\end{figure*}

Figure \ref{NminTestsEd} shows $\bar{N}_{min}$ as a function of the deposited energy for $\gamma=2$ obtained from simulations
following the procedure describe in Sec. \ref{Disc}.

From figure \ref{NminTestsEd} it can be seen that $\bar{N}_{min}(E)$ is larger than $N_{min}(E)$. The ratio 
$r(E)=\bar{N}_{min}(E)/N_{min}(E)$ can be approximated by a linear function of the logarithm of the deposited energy. Table 
\ref{RatioEdGamma} shows the variation of $r$ from $E_D=10^{19.2} \textrm{eV}$ to $E_D=10^{20.5} \textrm{eV}$ obtained form the 
linear fits for the different assumptions about the incident flux and for the corresponding tests. The results shown in Table 
\ref{RatioEdGamma} are obtained considering the region of $95\%$ probability for the true hypothesis, and similar results are 
obtained when the region of $68\%$ probability is considered. It can be seen that the discrimination between $1:1$ and $1:0$ is 
more affected by the use of the deposited energy instead of the true energy, which is due to the smaller values obtained for 
$\bar{R}_\nu(E)$ compared with the corresponding ones for $R_\nu(E)$. The discrimination between these two scenarios is very 
sensitive to $R_\nu$ and to $\bar{R}_\nu$ when the deposited energy is considered.      
\begin{table}[!ht]
\caption{\label{RatioEdGamma} $r(10^{19.2} \textrm{eV}) \rightarrow r(10^{20.5} \textrm{eV})$ for $\gamma=2$ and $\gamma=3$, obtained 
considering the region of $95\%$ probability for the true hypothesis.}
\begin{ruledtabular}
\begin{tabular}{cccc}
 True & Test & $\gamma=2$ & $\gamma=3$ \\ \hline
 $1:1$  & $1:0$  &  $3.1 \rightarrow 3.2$  &  $1.8 \rightarrow 2.4$  \\
 $1:1$  & $0:1$  &  $1.8 \rightarrow 1.2$  &  $1.2 \rightarrow 1.2$  \\
 $1:0$  & $1:1$  &  $3.2 \rightarrow 3.2$  &  $1.8 \rightarrow 2.4$  \\
 $1:0$  & $0:1$  &  $1.8 \rightarrow 1.4$  &  $1.2 \rightarrow 1.4$  \\
 $0:1$  & $1:1$  &  $1.9 \rightarrow 1.0$  &  $1.3 \rightarrow 1.2$  \\
 $0:1$  & $1:0$  &  $2.0 \rightarrow 1.0$  &  $1.4 \rightarrow 1.0$  \\
\end{tabular}
\end{ruledtabular}
\end{table}


\end{document}